\documentclass[12pt]{iopart}

\usepackage{graphicx}

\begin{document}

\topical[Deterministically fabricated solid-state quantum-light sources]{Deterministically fabricated solid-state quantum-light sources}


\author{Sven Rodt, Stephan Reitzenstein, and Tobias Heindel}

\address{Institute of Solid-State Physics, Technische Universit\"{a}t Berlin, Hardenbergstra\ss e 36, 10623 Berlin, Germany}
\ead{tobias.heindel@tu-berlin.de}
\vspace{10pt}
\begin{indented}
\item[]September 2018
\end{indented}

\begin{abstract}
The controlled generation of non-classical states of light is a challenging task at the heart of quantum optics. Aside from the mere spirit of science, the related research is strongly driven by applications in photonic quantum technologies, including the fields of quantum communication, quantum computation, and quantum metrology. In this context, the realization of integrated solid-state-based quantum-light sources is of particular interest, due to the prospects for scalability and device integration. 

This topical review focuses on solid-state quantum-light sources which are fabricated in a deterministic fashion. In this framework we cover quantum emitters represented by semiconductor quantum dots, colour centres in diamond, and defect-/strain-centres in two-dimensional materials. First, we introduce the topic of quantum-light sources and non-classical light generation for applications in photonic quantum technologies, motivating the need for the development of scalable device technologies to push the field to real-world applications. In the second part, we summarize material systems hosting quantum emitters in the solid-state. The third part reviews deterministic fabrication techniques and comparatively discusses their advantages and disadvantages. The techniques are classified in bottom-up approaches, exploiting the site-controlled positioning of the quantum emitters themselves, and top-down approaches, allowing for the precise alignment of photonic microstructures to pre-selected quantum emitters. Special emphasis is put on the progress achieved in the development of in-situ techniques, which significantly pushed the performance of quantum-light sources towards applications. Additionally we discuss hybrid approaches, exploiting pick-and-place techniques or wafer-bonding. The fourth part presents state-of-the-art quantum-dot quantum-light sources based on the fabrication techniques presented in the previous sections, which feature engineered functionality and enhanced photon collection efficiency. The article closes by highlighting recent applications of deterministic solid-state-based quantum-light sources in the fields of quantum communication, quantum computing, and quantum metrology, and discussing future perspectives in the field of solid-state quantum-light sources.
\end{abstract}
\clearpage
%
%
%
%
%

\tableofcontents
\clearpage

\section{Introduction}
Light sources capable for the generation of non-classical states of light are at the heart of many applications in photonic quantum technologies \cite{OBrien2009}. Many schemes of quantum communication and cryptography \cite{Gisin2002,Gisin2007,Lo2014,Flamini2018}, for instance, rely on the availability of streams of single photons or entangled photon pairs on demand. The desired quantum light states are thereby often required to meet specific properties regarding the photons' wavelength, coherence time, indistinguishability, or polarization. Examples include quantum key distribution \cite{Bennett1984} as well as quantum repeaters protocols \cite{Briegel1998,Sangouard2007}. Due to the prospects for device integration and scalability, engineered solid-state-based quantum-light sources are of particular interest. The respective quantum emitters, however, differ in their optoelectronic properties within and across given ensembles, due to statistical variations in size, geometry, and material composition or simply the solid-state hostmaterial itself. To engineer quantum devices with the required optical properties, deterministic fabrication techniques are in great demand, enabling a higher degree of control for the parameters relevant for their targeted applications.

This topical review focuses on solid-state quantum-light sources which are fabricated in a deterministic fashion using advanced nanotechnology platforms. In section \ref{sec:Mat} we first introduce different types of material systems hosting quantum emitters in the solid-state. Section \ref{sec:Fab} reviews the techniques used for deterministic device fabrication. The respective approaches are grouped in four categories: (1) Bottom-up techniques for the site-controlled growth or definition of quantum emitters (cf. section~\ref{sec:Bottom-Up}). (2) Top-down marker-based techniques for the device fabrication around pre-selected, self-organized and stochastically-grown quantum emitters (cf. section~\ref{sec:Top-Down}). (3) In-situ techniques, where the emitter selection or definition and the lithography of a device is achieved in the same machine at low temperatures (cf. section~\ref{sec:In-Situ}). (4) Hybrid approaches, combining different techniques from (1) to (3) (cf. section~\ref{sec:Hybrid}). In section \ref{sec:QuantumLightSources} state-of-the-art solid-state quantum-light sources based on the fabrication techniques presented in the previous sections are presented with a focus on quantum dot devices. Here, emphasis is put on an engineered functionality, enhanced photon collection efficiency, and recent developments such as the modularization in plug-and-play devices. The article closes with section \ref{sec:Applications} by highlighting recent applications of deterministic solid-state-based quantum-light sources in photonic quantum technologies and discussing future perspectives in this context.

\section{Quantum emitter and host materials}\label{sec:Mat}
Various types of material systems can be used for the deterministic fabrication of quantum-light sources. 
In this sections we introduce the most prominent types of quantum emitters and host materials for non-classical light generation in the solid-state.

\begin{figure}[t]
\begin{center}
\includegraphics[width=1.0\linewidth]{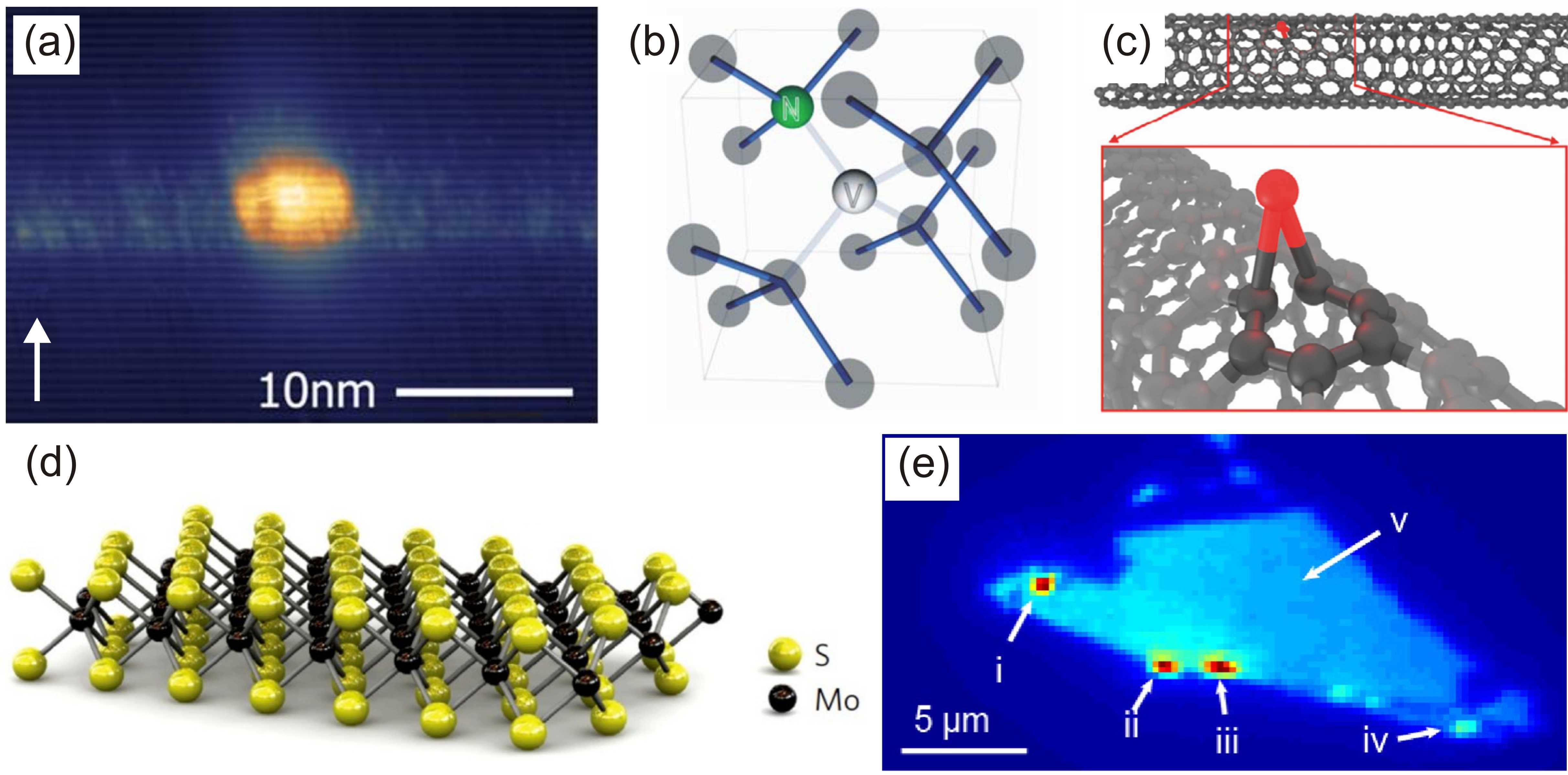}
\caption{Material systems used for the generation of non-classical light in the solid-state: (a) Self-organized InGaAs/GaAs QDs (adapted from \cite{Keizer2012}, with permission of AIP Publishing), (b) nitrogen-vacancies (NV) in diamond crystals (\cite{Jelezko2006}, reprinted with permission, \copyright 2006 WILEY-VCH), (c) Solitary dopants of carbon nanotubes (\cite{Ma2015a}, adapted by permission from Springer Nature, \copyright 2015 Macmillan Publishers Limited), and (d) defect centres in monolayers of MoS$_2$ (\cite{Radisavljevic2011},reprinted by permission from Springer Nature, \copyright 2011 Macmillan Publishers Limited) observed in a photolumincescence intensity map in (e) (reprinted with permission from \cite{Tonndorf2015}, \copyright The Optical Society.}
\label{fig:Fig_1}
\end{center}
\end{figure}

\subsection{Epitaxial semiconductor quantum dots}\label{Sec:QuantumDots}
Since their discovery in the early 1990s \cite{Leonard1993,Grundmann1995} epitaxial semiconductor quantum dots (QDs) have had an impressive career and rouse prospects for novel applications in various fields of research reaching from semiconductor laser physics \cite{Arakawa1982} to quantum information processing \cite{Imamoglu1999,Michler2009}.

QDs are crystalline clusters of a few hundreds to thousands of atoms embedded in a semiconductor matrix (see figure~\ref{fig:Fig_1}(a)). The QD islands can form in a self-organized fashion during the epitaxy of single monolayers of two materials with different lattice constants. Different types of growth regimes are distinguished depending on the materials and the growth conditions (temperature, pressure, amount of material, growth interruption, etc.) \cite{Lay1978}. Typically, the energy band gap of the QD material is chosen to be smaller as compared to the surrounding matrix material, which leads to a three-dimensional confinement potential for electrons and holes in the conduction band and valence band, respectively \cite{Petroff2001}. The growth of QDs can be conducted either via molecular beam epitaxy (MBE) or metal-organic chemical vapor deposition (MOCVD). Both methods enable monolayer deposition with high precision. In MOCVD reactant gases are fed into the reactor at a typical pressure of 15 to 750\,Torr, whereas MBE requires ultra-high vacuum conditions (pressures $<10^{-8}\,$Torr) during the epitaxy of the semiconductor material obtained from heated effusion cells \cite{Pohl2013}. While MBE is mostly used in research laboratories and enables lowest impurity levels, MOCVD is also employed for large scale production due to its lower costs.
The internal atomic structure of a QD embedded in bulk material is resolved by cross-sectional scanning tunneling microscopy (XSTM) in Figure~\ref{fig:Fig_1}\,(a) \cite{Keizer2012}.

The confinement potential of QDs with lateral dimensions on the order of the de Broglie wavelength of electrons and holes leads to quantized energy levels of the confined charge carriers and, hence, to a discrete emission spectrum in photo- or electroluminescence experiments. Especially at low temperatures, phonon-coupling is almost negligible~\cite{Borri2001,Besombes2001,Bayer2002}, which leads to predominant emission into the zero-phonon line (ZPL), in contrast to colour centres (cf. \ref{Sec:ColorCenters}). Besides the fundamental excitonic state X, constituted of a single electron and a single hole, various multiparticle states can form inside a single QD, typically leading to a variety of emission lines. Two electron-hole pairs captured inside a QD, for instance, form a biexciton (XX) state, while uneven numbers of charge-carriers for electrons and holes result in positively or negatively charged multi-particle states \cite{Gammon1996,Kulakovskii1999,Rodt2003,Sarkar2006,Seguin2006}.
Spectrally selecting the emission of one specific state results in the distillation of single photons leaving the QD one by one. This leads to the famous antibunching effect in photon statistics measurements via a Hanbury Brown and Twiss (HBT) setup~\cite{Brown1956}. Recently, a record-low antibunching value of $g^{(2)}(0)=(7.5\pm1.6)\times10^{-5}$ has been reported using QD single-photon sources \cite{Schweickert2018}. Beyond the possibility of generating single-photon states, QDs also allow for the generation of temporally-correlated photon pairs \cite{Regelman2001,Kiraz2002,Santori2002a} via the biexciton-exciton (XX-X) radiative cascade \cite{Benson2000}. As mentioned above, the XX state of a QD is constituted of two bound electron-hole pairs. Owing to quantum mechanical Coulomb interactions of the involved charge carriers \cite{Schliwa2009a}, this state typically shows a finite binding energy $E_{\rm{bin}}^{\rm{XX}}$ with respect to the case of two unbound excitons, which is on the order of 1\,meV in case of the InGaAs/GaAs material system \cite{Rodt2003}. The exciton state, on the other hand, consists of a single electron-hole pair and usually reveals a fine-structure splitting $\Delta E_{\rm{FSS}}$ on the order of 10\,$\mu$eV \cite{Gammon1996}, which arises from anisotropic electron-hole exchange interaction. The resulting radiative cascade emits pairs of photons in two possible decay channels, one being linear-horizontally (H) and the other one linear-vertically (V) polarized. Exploiting specific symmetry properties of QDs, this XX-X radiative cascade can produce polarization-entangled photon pairs \cite{Akopian2006,Young2006} or so-called twin-photon states \cite{Heindel2017,Moroni2019} (see section~\ref{sec:MultiPhoton}). For applications in photonic quantum technologies QDs are particular interesting, as they can be embedded straightforwardly in diode structures enabling electrical operation of the emitters \cite{Yuan2002,Salter2010}. Moreover, the radiative lifetime of QDs is relatively short ($\approx1\,$ns, Ref.~\cite{Aharonovich2016} and references therein), enabling high photon generation rates.

The stochastic nature of the self-organized growth of QDs, however, leads to an inhomogeneous broadening of the emission properties of ensembles of emitters. 
Therefore, various measures have been developed to influence the QD emission wavelength. Using partial capping and annealing during the QD growth \cite{Fafard1999,Wang2006b,Schneider2011} or post-growth rapid thermal annealing \cite{Young2005,Seguin2006,Ellis2007a,Dousse2010} the emission wavelength and other properties, such as the fine-structure splitting, of QDs can be adjusted to a certain extent at the ensemble level. Still these techniques are only workarounds towards fully deterministic and scalable technology platforms, which will be the central subject of this review article.

\subsection{Colour centres in crystals}\label{Sec:ColorCenters}
Another promising type of quantum emitters are colour or defect centres in bulk or nano crystals.
colour centres are crystallographic point defects, where a single native atom of the lattice is substituted by an impurity (see figure~\ref{fig:Fig_1}(b)).
This configuration leads to discrete electronic states deep inside the energy bandgap of the host crystal, often resulting in a high temperature stability up to room temperature \cite{Kurtsiefer2000}.
Prominent examples are nitrogen-vacancies (NV) or silicon-vacancies (SiV) in diamond crystals, which are to date among the most thoroughly studied colour centres (see Refs.~\cite{Jelezko2006,Aharonovich2014} for a review).
NV and SiV centres naturally occur in diamond, but can also be produced deterministically, which will be discussed in section \ref{sec:In-Situ}. 
Compared to QDs, the coupling to phonons is relatively strong. The SiV centre in diamond, however, also provides strong emission into the zero phonon line (ZPL) of about 70\% at room temperature and additionally enables the fabrication of emitter ensembles with ultra-small inhomogeneous broadening \cite{Rogers2014}. More recently also germanium-vacancy (GeV) or tin-vacancy (SnV) color centres in diamond are considered to enable single-photon sources with improved quantum efficiencies \cite{Iwasaki2015}.
Compared to QDs, the radiative lifetimes of defect centres in diamond are long-lived ($\approx10\,$ns, Ref.~\cite{Aharonovich2016} and references therein), leading to reduced photon generation rates.

Beyond the diamond material platform, large progress has been achieved with colour centres in compound semiconductors. 
Silicon carbide (SiC) with embedded positively charged carbon antisite-vacancies (C$_{\rm{V}}$C$_{\rm{Si}}$) can be used to develop room-temperature single-photon sources \cite{Castelletto2013}.
Another wide-bandgap compound semiconductor with potential for the engineering of quantum light emitting devices is ZnO. With respect to the hosted colour centres, however, this material system is far less understood and the quantum emitters suffer from reduced optical quality as compared to the material systems discussed above \cite{Choi2014}. Additionally, the fabrication of high-quality diode structures has not been successful to date, hindering the realization of electrically controlled devices so far \cite{Aharonovich2016}.
Other candidates for single-photon generation in the solid-state are rare-earth-ion impurities in crystals such as yttrium aluminum garnet (YAG) and yttrium orthosilicate (YOS) \cite{Kolesov2012,Eichhammer2015}.
Radiative lifetimes of colour centres in compound semiconductors (SiC, ZnO, etc.) and rare-earth impurities in YAG crystals are in between those of QDs and colour centres in diamond ($\approx1-4\,$ns, Ref.~\cite{Aharonovich2016} and references therein).

More generally, the realization of integrated devices based on colour centres is very demanding, and electrical pumping of the quantum emitters has been impossible for a long time.
For the diamond and SiC material system, this challenge has been mastered meanwhile, and single-photon emitting diodes operating at room-temperature were demonstrated \cite{Mizuochi2012,Lohrmann2015}.

\subsection{Two-dimensional materials}\label{sec:2D}
There exists a plethora of quasi two-dimensional (2D) materials, which can be obtained by exfoliation from bulk materials that are composed of stacks of weakly interacting atomically thin layers. The most prominent example for such a van-der-Waals material comprises a sheet of carbon atoms aligned in a hexagonal lattice - also known as graphene. 'Rolled up' forming a carbon nanotube (see figure~\ref{fig:Fig_1}(c)), graphene has been used to demonstrate antibunching in photoluminescence experiments \cite{Hoegele2008}. Very recently, carefully synthesized triangular flakes of graphene, referred to as graphene QDs, have been used for single-photon generation at room temperature \cite{Zhao2018}.

With respect to quantum light generation, recently also another type of 2D material system attracted great interest: Transition metal dichalcogenides (TMDCs) \cite{Wang2012,Mak2016}. TMDCs are a class of materials with the formula MX$_2$, where M is a transition metal element (group IV, V or VI), and X is a chalcogen (see figure~\ref{fig:Fig_1}(d)). One of the intriguing properties of TMDCs are their layer-dependent optical properties. Several TMDC semiconductors, for instance molybdenum disulfide (MoS$_2$) and tungsten diselenide (WSe$_2$), show a transition from an indirect bandgap in bulk to a direct bandgap in the monolayer configuration \cite{Mak2010,Splendiani2010}. In 2015 five groups independently reported the observation of single-photon emission by localized luminescence centres in WSe$_2$ \cite{He2015,Srivastava2015,Chakraborty2015,Koperski2015,Tonndorf2015}. Moreover, signatures of the XX-X radiative cascade were observed \cite{You2015,He2016a}, rising prospects for the generation of entangled photon pairs in 2D material systems. The origin of these emitters is attributed to excitons bound by shallow confinement potentials generated by local strain fluctuations. For this reason, TMDC-based quantum emitters to date operate only at cryogenic temperatures and are highly susceptible to spectral diffusion \cite{Tonndorf2015}.

In contrast, quantum emitters in hexagonal boron nitride (hBN) offer superior temperature stability, as the associated defect states are located deep inside the energy bandgap, similarly to colour centres in diamond. Therefore, single-photon emission is stable up to room temperature \cite{Tran2015,Tran2016} and even far beyond \cite{Kianinia2017}.

Radiative lifetimes of quantum emitters in 2D materials are similar to those of QDs (cf. Ref. \cite{Aharonovich2016} and references therein).
The emission, however, often suffers from pronounced spectral diffusion, which complicates quantum optics experiments. To date, for example, there is only a single report on resonant laser excitation spectroscopy of quantum emitters in non-deterministically fabricated 2D materials \cite{Kumar2016}. The difficulty of resonantly exciting quantum emitters in 2D materials in turn might be one reason why the generation of indistinguishable photons has not been demonstrated for this material system yet.

\section{Deterministic fabrication techniques}\label{sec:Fab}
For the scalable fabrication of quantum devices with high yield and optimal performance, deterministic fabrication techniques are required. The development and the ongoing improvement of such techniques is therefore one major driving force in the field of photonic quantum technologies. In this section, we introduce prominent approaches for embedding single quantum emitters in photonic devices using bottom-up or top-down techniques, where emphasize is put on recent technological developments. In section \ref{sec:Bottom-Up} the site-controlled growth or definition of quantum emitters will be reviewed. Marker-based approaches for deterministic device fabrication are discussed in section \ref{sec:Top-Down}). Section \ref{sec:In-Situ} focuses on in-situ techniques - approaches which are particularly in vogue today. Last but not least, section \ref{sec:Hybrid} presents hybrid deterministic approaches based on multiple material systems.

\subsection{Site-controlled quantum emitters (bottom-up)}\label{sec:Bottom-Up}
\begin{figure}[t]
\begin{center}
\includegraphics[width=1.0\linewidth]{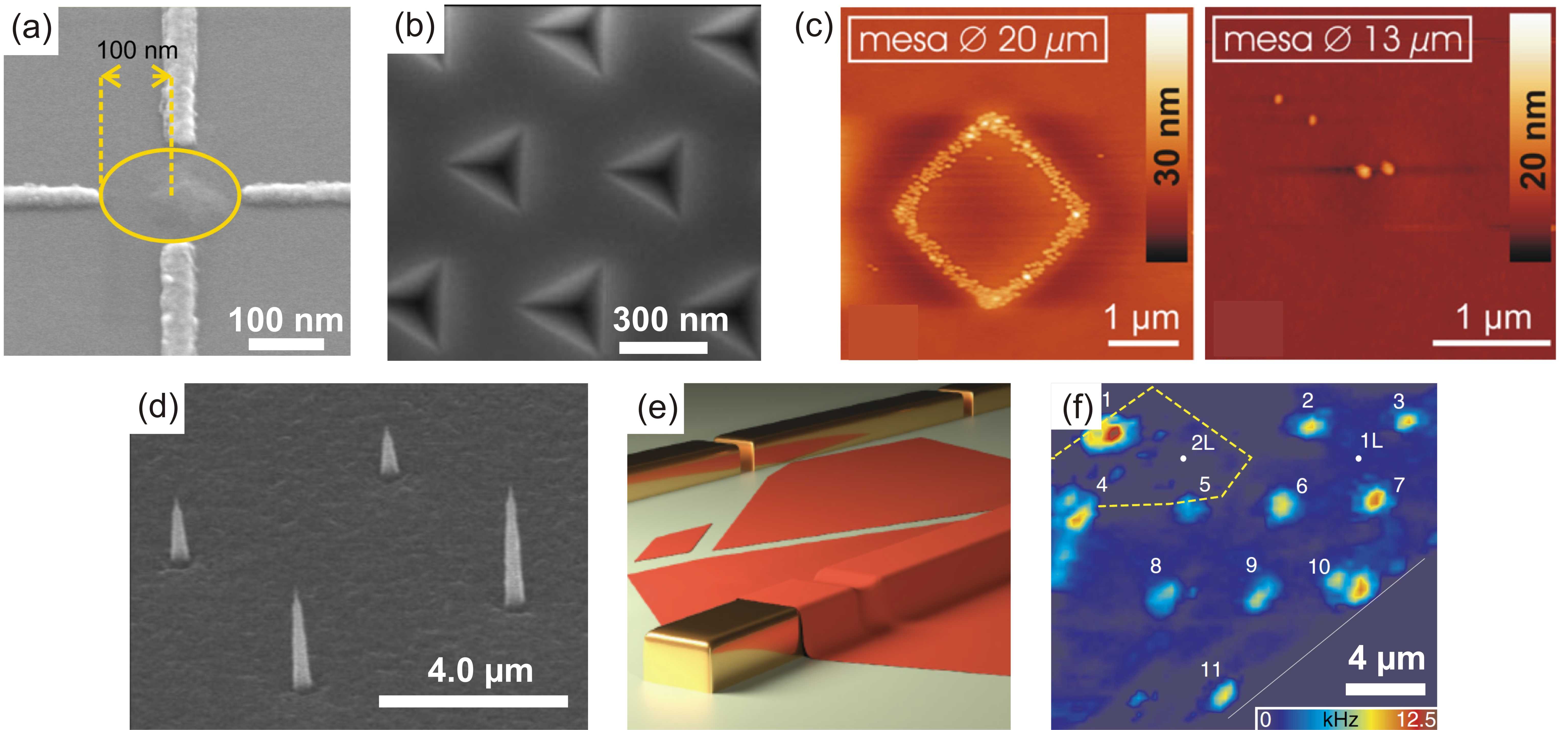}
\caption{Bottom-up techniques employed for the spatially deterministic positioning of single quantum emitters: (a) Site-controlled growth of InGaAs/GaAs QDs on etched nanoholes (reprinted from \cite{Schneider2008}, with permission of AIP Publishing), (b) pyramidal QDs based on the overgrowth of tetrahedral pits (\cite{Felici2009}, reprinted with permission, \copyright 2009 Wiley-VCH), (c) QD positioning via strain induced by a buried oxide aperture (adapted from \cite{Strittmatter2012}, with permission of AIP Publishing), (d) positioned nanowires with integrated QDs (adapted from \cite{Heinrich2010}, with permission of AIP Publishing), and strain-induced defect centres in 2D materials on (e) gold nanorods (\cite{Kern2016}, reprinted with permission, \copyright 2016 Wiley-VCH) and (f) dielectric nanopillars (\cite{Branny2017}, CC BY 4.0).}
\label{fig:Fig_2}
\end{center}
\end{figure}

To achieve the highest degree of scalability in quantum device fabrication, ideally the quantum emitter itself needs to be positioned precisely at a pre-defined location. In this section we will summarize such bottom-up approaches for QDs as well as for emerging 2D materials (see figure~\ref{fig:Fig_2}).
One method for achieving position-control for single QDs is the deposition of QD material on a pre-patterned substrate with or without an additional buffer layer \cite{Schmidt2007}. Using electron-beam lithography (EBL) in combination with dry etching, pits or nano-holes can be defined in the substrate. Overgrowth of this patterned substrate with QD material leads to a site-selective growth of QDs. As an prominent example for this approach Hartmann et al. demonstrated the growth of GaAs/AlGaAs heterostructures in inverted tetrahedral pyramids \cite{Hartmann1998,Felici2009}. In this case, hexagonal arrays of pits are etched into undoped (111)-B oriented GaAs substrate. In a next step Al$_{0.45}$Ga$_{0.55}$As/GaAs/Al$_{0.45}$Ga$_{0.55}$As single quantum well layer sequences are deposited via MOCVD. This growth configuration leads to the formation of inverted pyramids, where quantum wires and quantum wells develop at the pyramid corners and facets, while a single GaAs QD is formed at the sharp pyramid tip. The resulting QDs feature a precise position control ($<10\,$nm), small inhomogeneous broadening of the ensemble emission ($<10\,$meV) as compared to standard self-organized QDs (typically 30-60\,meV \cite{Leonard1993,Grundmann1995}), and can be adjusted in their emission wavelength \cite{Felici2009}. Due to the high degree of symmetry present in this approach, site-controlled pyramidal QDs show small fine-structure splittings $<20\,\mu$eV \cite{Mereni2012}, which has been exploited for the generation of polarization-entangled photon pairs (cf. \ref{sec:EL} and \ref{sec:MultiPhoton}). Another example to achieve site-controlled growth of QDs by using pre-patterened substrates was demonstrated by Schneider et al. \cite{Schneider2008}. Here, circular nano-holes with a diameter of 30\,nm are defined on (100) oriented GaAs substrate using electron-cyclotron-resonance reactive-ion or wet-chemical etching. In a next step, the nanohole surface is smoothed by a 12\,nm thick GaAs layer before the growth of InAs QDs and a 50\,nm thick capping layer is performed using MBE. Using this process, a standard deviation of the QD position of 50\,nm relative to the target location has been reported. This approach can be used to produce ordered arrays of single QDs with pitches between 200\,nm and 10$\,\mu$m \cite{Verma2011,Huggenberger2011a}, being beneficial for device integration. Introducing one or multiple separation layers between the nano-holes and the QD layer, the optical quality of this type of site-controlled QDs can be improved \cite{Schneider2012a}, which otherwise suffer from large spectral linewidth due to spectral diffusion caused by the nearby etched surfaces \cite{Albert2010}. Employing such refined approaches, J\"ons et al. demonstrated the triggered generation of indistinguishable photons emitted by site-controlled QDs \cite{Joens2013}.

Pre-patterned substrates can also be used to achieve site-controlled growth of photonic nanowires with integrated single quantum emitters. In their work, Heinrich et al. demonstrated the positioned growth of AlGaAs nanowires containing an axial GaAs QD using solid-source MBE in the vapor liquid solid mode \cite{Heinrich2010}. This approach resulted in tapered nanowires with an average diameter of 167\,nm and 304\,nm at the top and the bottom of the nanowire, respectively, and an average nanowire length of 2.6\,$\mu$m. A spectral linewidth of 95\,$\mu$eV and photon antibunching were observed with moderate single-photon purity ($g^{(2)}(0)=0.46$) due to uncorrelated background
emission from the doped GaAs substrate.

Another method for achieving high degrees of position control is based on a buried stressor consisting of an oxide aperture. Strittmatter et al. demonstrated, that buried oxide apertures generate a strain field at the surface of a GaAs buffer layer, which leads to the nucleation of QDs above the edges of the aperture. Reducing the aperture size, the strain-field can be ideally focused down to a small region, leading to the strain induced nucleation of a single site-controlled QD \cite{Strittmatter2012,Strittmatter2012a}. A particular appealing feature of this approach is, that the position control of a quantum emitter can be combined with an optical as well as an electrical confinement generated via the oxide aperture \cite{Ellis2007}, which is beneficial for the development of efficient deterministically-fabricated and electrically-driven devices (see section \ref{sec:EL}).

A completely different approach is used for achieving site-control of quantum emitters in 2D materials. In two-dimensional sheets of material exfoliated on planar surfaces, spatially localized quantum emitters with photon antibunching have been observed (cf. section~\ref{sec:2D}). The origin of the underlying emission process was still under debate, but the associated localized luminescence occurred mostly at the edges of flakes and could also be induced by scratching the material \cite{Tonndorf2015}. In 2015 Kumar et al. showed, that localization of quantum emitters can be generated by straining mono- and bilayer WSe$_2$ by patterned substrates \cite{Kumar2015}. In another approach Kern et al. demonstrated the nanoscale positioning of quantum emitters in a WSe$_2$ flake, covering a nanoscale gap between two gold nanorods thus resulting in a strain field \cite{Kern2016}. Shortly after, the realization of ordered arrays of quantum emitters was reported using WSe$_2$ flakes covering a substrate with nanopillars \cite{Branny2017,Palacios-Berraquero2017}. Additionally, also nanobubbles have been used to induce quantum emitters in WSe$_2$ and BN/WSe$_2$ heterostructures \cite{Shepard2017}.

The bottom up approaches presented in this section proved their potential for the scalable fabrication of quantum-light sources. The quantum optical properties of the respective QDs, however, often do not reach the excellent level of stochastically-grown self-organized QDs. One possible reason for the reduced optical quality is that the subjacent patterned substrate used to achieve site-control introduces defects close to the QDs. Additionally, the degree of control achievable for the emitter properties (e.g. emission wavelength, inhomogeneous broadening, etc.) to date is still not sufficient for many applications in quantum information technologies. These drawbacks are circumvented by top-down approaches presented in the following section.
These drawbacks are circumvented by top-down approaches presented in the following section.

\subsection{Marker-based approaches (top-down)}\label{sec:Top-Down}
\begin{figure}[t]
\begin{center}
\includegraphics[width=0.5\linewidth]{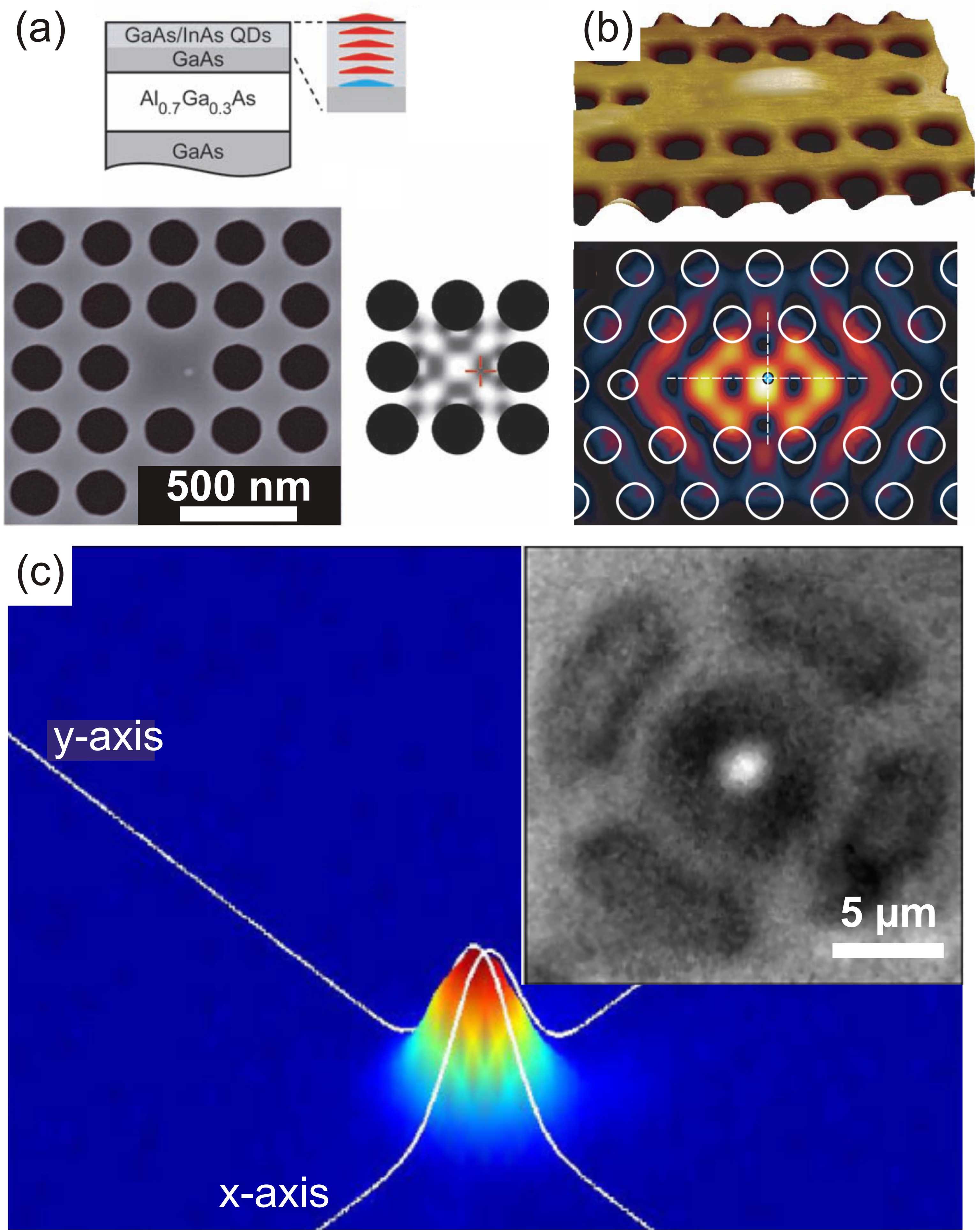}
\caption{Top-down marker-based approaches used for the deterministic fabrication of photonic microstructures with a single quantum emitter: (a) (from \cite{Badolato2005}, reprinted with permission from AAAS) and (b) (\cite{Hennessy2007}, reprinted by permission from Springer Nature, \copyright 2007 Nature Publishing Group) InAs/GaAs QDs inside a photonic crystal cavity (PhC), and (c) QDs integrated in circular Bragg gratings using nanoscale optical positioning (\cite{Sapienza2015}, CC BY 4.0).}
\label{fig:Fig_3}
\end{center}
\end{figure}

In top-down marker-based approaches, specific quantum emitters are pre-selected on the basis of their key emission properties and subsequently integrated into a photonic device. During a first step, suitable quantum emitters are located precisely relative to the markers defined onto the surface of the target sample. In the second step, the photonic structure is defined lithographically directly at the position of the quantum emitter. Seminal experiments using marker-based deterministic technologies are summarized in Figure~\ref{fig:Fig_3}.
In 2005 Badolato et al. demonstrated the deterministic coupling of a stochastically grown InAs/GaAs QD to the cavity mode of a photonic crystal cavity (PhC) with a Q-factor of about 3000 \cite{Badolato2005}. To be able to spatially detect the target QD buried in the unprocessed sample, six vertically strain-correlated QD layers were grown above the spectrally blue-detuned target quantum emitter. The uppermost QD could be resolved in SEM images relative to gold markers, which was then used to define the spatially aligned PhC via EBL. This deterministic device enabled the observation of the Purcell effect \cite{Purcell1946,Gerard1998} in the weak coupling regime. A refined method of the same group enabled two years later the observation of a deterministically-fabricated strongly-coupled QD-cavity system \cite{Hennessy2007}. This was achieved by fabricating a high-Q (13,300) PhC around a single QD using marker structures yielding a spatial accuracy of 30\,nm.

In the work discussed above, the localization of QDs was performed via SEM or AFM, respectively, while the optical properties had to be checked in another characterization step via micro-photoluminescence ($\mu$PL) spectroscopy before fabricating the final device. The prospects and limitations of such combined methods were studied in detail in Ref. \cite{Sapienza2017}. In contrast, the reports presented in the following enable one to localize and spectrally analyze the target QDs in a single imaging step, while device fabrication was still performed by marker-based EBL. In Ref.~\cite{Nogues2013} CL spectroscopy was used to locate single QDs with respect to markers and plasmonic nano-antennas were fabricated by EBL with an alignment accuracy better than $50\,$nm. A similar lateral accuracy was reached in Ref.~\cite{Kojima2013} by combining $\mu$PL and EBL with the aid of markers to integrate single QDs into photonic crystals. Also, marker-based $\mu$PL together with EBL was used in Ref.~\cite{Sapienza2015} to realize circular Bragg gratings around single QDs with an accuracy of better than 30~nm. More recently, the same group reported a further improved positioning uncertainty of $4.5\,$nm using a technically refined method \cite{Liu2017}.

A potential drawback of the marker-based deterministic approaches presented above is the two-step process. While the optical pre-selection is performed for instance in a $\mu$PL-setup, the sample processing is conducted via photolithography or EBL in a dedicated lithography machine. The spatial correspondence between the two setups needs to be assured using marker structures on the sample surface that were fabricated beforehand. These two steps make the fabrication more complex and therefore more susceptible with respect to misalignments. The approaches presented in the following perform both steps in a single machine, and therefore potentially offer increased alignment accuracy while reducing the process complexity.

\subsection{In-situ techniques}\label{sec:In-Situ}
A particular powerful class of approaches used for deterministic device fabrication comprises {\it in-situ} techniques. Here, the pre-selection or creation of quantum emitters and, if applicable, also the lithographic definition of a surrounding device are performed in one and the same apparatus without transferring the sample or heating it up (if the pre-selection was performed at cryogenic temperatures) \cite{Donatini2010}. This has several appealing prospects: Firstly, the alignment accuracy is improved, as the coordinate system is identical for both steps. Secondly it speeds up the fabrication, as no marker or coordinate-system matching is required for each write field. Not least, the setup complexity is strongly reduced, as only a single machine is necessary.

\begin{figure}[t]
\begin{center}
\includegraphics[width=0.5\linewidth]{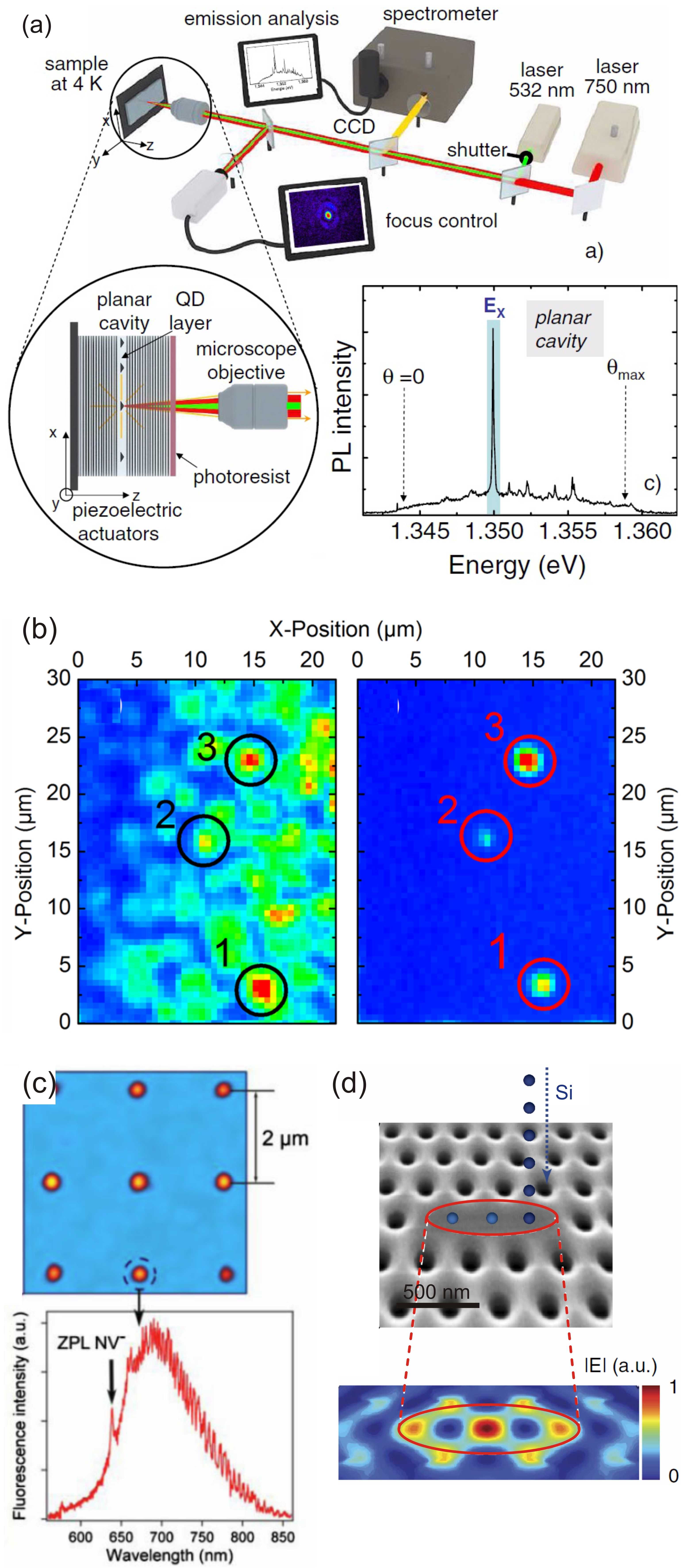}
\caption{Top-down in-situ lithography techniques for deterministic device fabrication: (a) In-situ optical lithography of QD-micropillar cavities (\cite{Dousse2008}, reprinted figure with permission, \copyright 2008 by the American Physical Society), (b) in-situ EBL of QD-mesas (reprinted from \cite{Gschrey2013}, with permission of AIP Publishing), and ion-implantation of arrays of (c) NV centres (\cite{Lesik2013}, reprinted with permission, \copyright 2013 Wiley-VCH) and (d) SiV centres (\cite{Schroeder2017}, CC BY 4.0) in diamond.}
\label{fig:Fig_4}
\end{center}
\end{figure}

A prominent example based on in-situ optical lithography is the work by Dousse et. al \cite{Dousse2008} (see figure~\ref{fig:Fig_4}\,(a)). This technique is based on a $\mu$PL setup using two different lasers with emission wavelengths of 750\,nm and 532\,nm, respectively. The sample is coated with a photo-sensitive resist before it is mounted inside a cryostat for the cool-down to low temperatures (10\,K). Using the 750\,nm laser, the resist is not exposed and a suitable target QD can be located using $\mu$PL-spectroscopy. After QD selection, the green laser is used to expose the resist right above the quantum emitter. In the next steps, the resist is developed at room temperature and a lift-off process together with chloride reactive ion etching is applied to define a micropillar cavity containing the pre-selected target QD. This approach enables a spatial accuracy for the positioning of micropillars of 50\,nm. The fabrication of bright quantum-light sources using in-situ optical lithography will be discussed in section \ref{sec:SPS} and \ref{sec:MultiPhoton}.

Inspired by the work presented above, a deterministic device fabrication technique based on in-situ EBL has been developed. In their work, Gschrey et al. combined EBL with cathodoluminescence spectroscopy \cite{Gschrey2013}. For this purpose, a scanning-electron microscope (SEM) was equipped with an extension for cathodoluminescence lithography \cite{Yacobi1986}. Here, the sample is first spin-coated with an electron-beam resist at room-temperature. After cooling down the sample to cryogenic temperatures of about 10\,K, pre-selection of target QDs based on their brightness and emission wavelength is performed by scanning the quantum emitters with the electron beam and mapping the luminescence using a spectrometer attached to the SEM. In the next step, the electron beam is used to define a photonic structure in the resist. As the resist is already exposed during the pre-selection process, the writing of the final structure is performed by inverting the resist, which directly acts as etch mask for the final etch step after resist development at room-temperature. Advantages of in-situ EBL, compared to the optical counterpart, include (at least in principle) improved accuracy as well as high resolution, and, in case of more complex structures such as waveguide circuits, also higher speed. The overall lateral accuracy has been reported to be 34\,nm \cite{Gschrey2015}, mainly limited by temperature-induced mechanical drifts of the sample holder inside the SEM. A spatial resolution below 10\,nm should be within reach based on state-of-the-art EBL systems. The speed for writing extended device patterns in the resist is faster compared to optical in-situ lithography, as the electron-beam can be scanned quickly across the sample. In Ref.~\cite{Gschrey2013}, this approach has been used to fabricate deterministic QD mesa structures with high device yield and high quantum optical properties. In section \ref{sec:SPS} and \ref{sec:MultiPhoton} the fabrication of deterministically fabricated QD quantum-light sources will be presented, which are based on a refined in-situ EBL technique.

The in-situ lithography approaches presented above have so far been employed mostly for QD systems. For colour centres, ion-implantation has been proven to be a powerfull technique for the deterministic emitter positioning, which can be combined with in-situ processing of photonic devices. In Ref.~\cite{Meijer2005} Meijer et al. demonstrated the generation of arrays of NV centres in a diamond crystal by implanting nitrogen atoms and annealing the sample. For this purpose a beam of N$^+$ ions produced by a dynamitron tandem accelerator with an energy of 2\,MeV and a diameter of 0.3\,$\mu$m was scanned stepwise across the sample. Later on, the implantation of ions for the generation of NV centres has been implemented using focused ion beam (FIB) technology in combination with a SEM in a dual-beam configuration \cite{Lesik2013} (see figure~\ref{fig:Fig_4}\,(c)), which is much more practical then using a tandem accelerator. This technique has then also been adapted for the generation of SiV centres in diamond \cite{Tamura2014,Schroeder2017} (see figure~\ref{fig:Fig_4}\,(d)) as well as silicon carbide \cite{Radulaski2017} crystals, and more recently also germanium-vacancies (GeV) in diamond \cite{Zhou2018}. By combining ion-implantation with plasma etching techniques, photonic microstructures such as nanopillars \cite{Radulaski2017} or photonic-crystal cavities and waveguides \cite{Schukraft2016} can be realized in-situ. These approaches, however, doo not allow for the controlled implantation of a single ion with 100\% yield, leading to a Poissonian distribution of the numbers of NV centres per site.

\subsection{Hybrid approaches}\label{sec:Hybrid}
\begin{figure}[t]
\begin{center}
\includegraphics[width=1.0\linewidth]{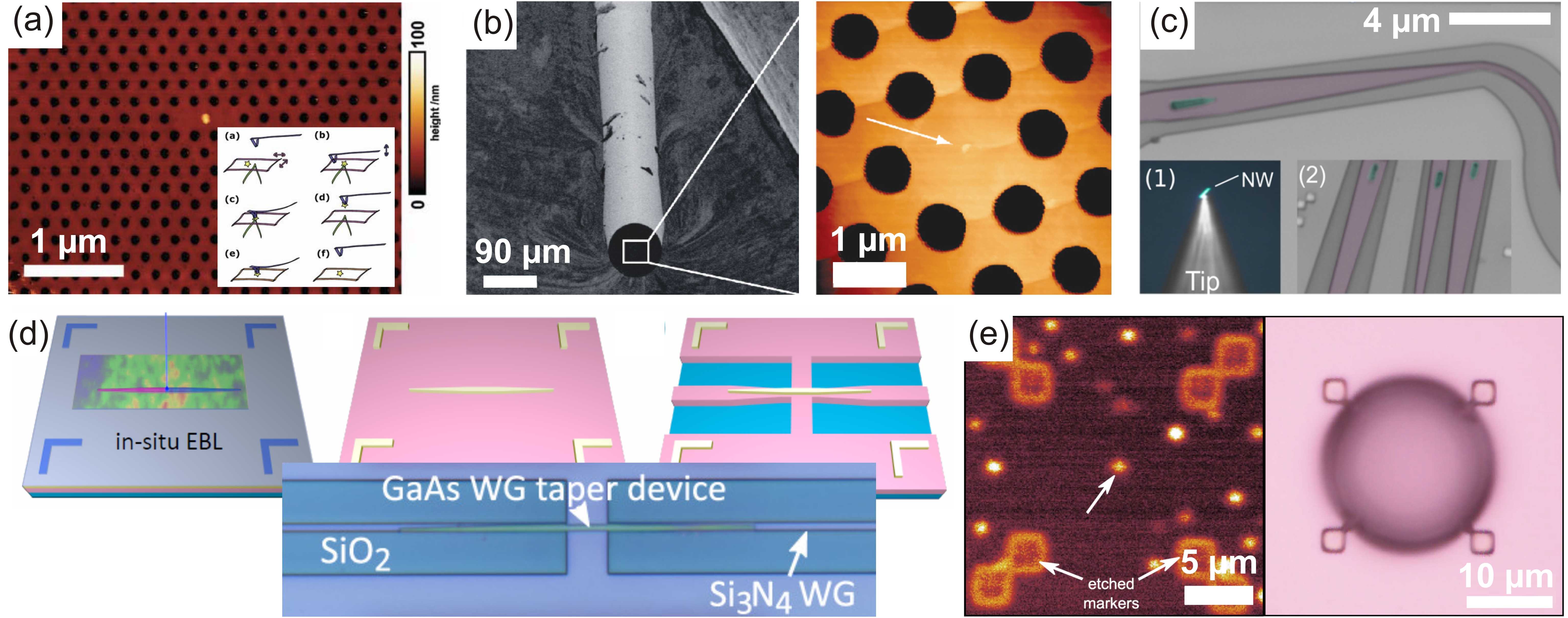}
\caption{Hybrid approaches for the deterministic positioning of quantum emitters in photonic devices: Nano-diamonds (a) inside a PhC (adapted from \cite{Schell2011} with permission, \copyright 2011 American Institute of Physics) and (b) at the facet of a photonic crytstal fiber (\cite{Schroeder2011}, reprinted with permission, \copyright 2011 American Chemical Society). (c) Nanowire-QD integrated into a SiN waveguide (\cite{Zadeh2016}, adapted with permission, \copyright 2016 American Chemical Society). (d) Pre-selected QD integrated into a tapered GaAs waveguide coupled to a SiN waveguide (\cite{Schnauber2019}, adapted with permission, \copyright 2019 American Chemical Society.). (e) Solid immersion lens (SIL) fabricated by laser writing above a pre-selected QD (\cite{Sartison2017}, CC BY 4.0).}
\label{fig:Fig_5}
\end{center}
\end{figure}

In some cases, it is beneficial to combine different material systems for the quantum emitter on one hand and the photon collection or guiding on the other hand, to gain synergies of both worlds.
Such device approaches in many cases need to make use of transfer techniques to integrate the active nanophotonic part (including the quantum emitter) with the passive part for enhanced photon collection and guiding (waveguides, cavities or solid immersion lenses). Examples for transfer techniques include nano-manipulation using the tip of an atomic force microscope (AFM) \cite{Junno1995,Toset2006} or a scanning electron microscope (SEM) \cite{Sar2009,Ampem-Lassen2009} or wafer-bonding \cite{Davanco2017}, or so-called transfer printing, where the emitter is embedded in a thin film of rubber \cite{Katsumi2018} or polymer \cite{Reimer2012}. Hybrid devices fabricated in a deterministic fashion are summarized in Figure~\ref{fig:Fig_5}. In Ref.\cite{Schell2011}, the tip of an AFM has been used for picking, transferring and placing single nanodiamonds containing NV centres. Applying forces of 1$\,\mu$N was sufficient to attach the nanodiamond to the tip via surface adhesion. To monitor the success of the pick-up procedure, the fluorescence of the nanodiamond was observed during tip movement. In that way, nanodiamonds can be placed at the center of a gallium phosphide photonic crystal membrane cavity (see figure~\ref{fig:Fig_5}\,(a)). The yield for this procedure was reported to be about one third. Embedding the quantum emitter in a photonic device, can improve the photon collection efficiency. Placing a nanodiamond for instance inside a PhC membrane cavity, Wolters et al. demonstrated the Purcell enhancement of the ZPL emission of a single NV center \cite{Wolters2010}. For the development of practical devices, also the coupling to optical fibers is highly desirable. This has been demonstrated for NV centres in nanodiamonds positioned at the facet of a photonic crystal fiber \cite{Schroeder2011} (see figure~\ref{fig:Fig_5}\,(b)) or the waist of a tapered optical fiber \cite{Liebermeister2014} using the pick-and-place technique discussed above. A fiber-coupled single-photon emitting device based on a similiar approach adapted for nanowires will be discussed in section \ref{sec:PnP}. Furthermore, nanowires with integrated single QDs have also been deterministically integrated in silicon-based photonic waveguides combining pick-and-place and waveguide processing \cite{Zadeh2016} (see figure~\ref{fig:Fig_5}\,(c)). Moreover, the deterministic fabrication of hybrid GaAs/Si$_3$N$_4$ waveguide devices with integrated pre-selected QDs (see figure~\ref{fig:Fig_5}\,(d)) was recently demonstrated by Schnauber et al. in an all-lithography-based approach \cite{Schnauber2019}, i.e. without requiring pick-and-place. For this purpose the authors combined in-situ EBL on GaAs-based QD devices \cite{Gschrey2013} with waferbonding to silicon-based waveguides \cite{Davanco2017}. Finally, Sartison et al. combined in-situ photolithography with 3D laser writing to first preselect a single QD, define markers, and then fabricate a solid immersion lens on top of it \cite{Sartison2017} (see figure~\ref{fig:Fig_5}\,(e)).\\

\section{Deterministic solid-state quantum-light sources}\label{sec:QuantumLightSources}
Enormous efforts are being made for the development of efficient devices for quantum light generation, being one of the most demanding building blocks for applications in photonic quantum technologies.
In this section we review solid-state-based quantum-light sources which are fabricated deterministically by using the approaches presented in the previous section.

The performance of the discussed quantum-light sources will be compared using the following quantities: The quality factor, the photon extraction efficiency, the antibunching value, the photon indistinguishability, and the entanglement fidelity. In the following we briefly introduce each quantity.

The \textbf{quality (Q) factor} of an optical cavity is a measure for the storage time of photons inside the resonator. Experimentally, the Q-factor is extracted from spectoscopic measurements and is given by the resonator's frequency-to-bandwidth ratio $Q=\omega/\delta\omega$.

The \textbf{photon extraction efficiency} refers to the fraction of the total photon flux emitted by an emitter which is collected by a certain numerical aperture (NA) defined by the experimental apparatus. Depending on the far-field characteristics of a photonic device, smaller or larger NAs are sufficient to achieve high photon extraction efficiencies into the 'first lens'. The photon extraction efficiency can be determined by a careful calibration of the experimental setup \cite{Strauf2006,Claudon2010}.

The \textbf{antibunching value $g^{(2)}(0)$} is a measure for the probability that a light sources emits two photons at the same time and refers to the case of $\tau=t_{\rm{2}}-t_{\rm{1}}=0$ (zero temporal delay) for the photon autocorrelation $g^{(2)}(\tau)$ of a given quantized light field \cite{Glauber1963}. An ideal single-photon source emits exactly one photon at a time leading to $g^{(2)}(0)=0$, while $g^{(2)}(\tau)=1$ for other delay times (sub-Poissonian statistics). A perfectly coherent classical light source, e.g. a laser, is characterized by $g^{(2)}(0)=1$ (Poissonian statistics), while classical thermal light fields, emitted e.g. by light bulbs or fluorescent lamps, show $1<g^{(2)}(0)<2$ (super-Poissonian statistics). Experimentally, $g^{(2)}(0)$ is determined via photon autocorrelation measurements in a Hanbury-Brown and Twiss type setup \cite{Brown1956}. In experiments under pulsed excitation, $g^{(2)}(0)$ is evaluated as the integrated area of the $\tau=0$ coincidence peak divided by the average area of the coincidence peaks at finite $\tau$. Note, that $g^{(2)}(0)$ does neither depend on the vacuum contributions present in the light field, i.e. the efficiency of the light source, nor on the decoherence.

The \textbf{photon indistinguishability} corresponds to the mean wavefunction overlap of two photons from a statistical ensemble. Photons are called fully indistinguishable, if they can be described by the same set of identical quantum numbers. The photon indistinguishability can be determined via two-photon interference experiments in a Hong-Ou-Mandel (HOM) type setup \cite{Hong1987,Santori2002b}. Here, two photons enter a 50:50 beamsplitter from different ports and interfere with each other. In case of perfect indistinguishabillity, both photons always leave the beamsplitter in the same but stochastically random exit port, as the probability amplitudes destructively interfere for the cases where both photons are transmitted or reflected. This HOM effect can be observed as antibunching in coincidence measurements at both exit ports. In contrast to the antibunching in photon autocorrelation measurements, the photon indistinguishability is crucially affected by decoherence. Typically two measurements are performed to experimentally determine the photon indistinguishability, one where the input photons have parallel polarization orientation and the other one with orthogonal polarization configuration. The contrast or visibility between both measurements reveals the photon indistinguishability. While most experiments use binary click detectors for HOM experiments, photon-number resolving detectors are an interesting alternative \cite{DiGiuseppe2003,Helversen2019}.

The \textbf{entanglement fidelity $F^{+}$}, as referred to in this review article,  is defined as the overlap of an experimentally generated two-photon wavefunction $\left|\Phi\right\rangle$ with the maximally entangled Bell state $\left|\Psi^+\right\rangle=1/\sqrt{2}(\left|H_{\rm{XX}}H_{\rm{X}}+V_{\rm{XX}}V_{\rm{X}}\right\rangle)$. Note, that this Bell state is entangled in the polarizations degree of freedom, while entanglement can also be realized in different degrees of freedom, also simultaneously \cite{Prilmueller2018}. Experimentally, the entanglement fidelity is determined using quantum tomography \cite{James2001,Akopian2006}, where $F^{+}=0.5$ corresponds to a perfectly polarization-correlated but classical state and $F^{+}=1$ to a maximally entangled state. Note, that also other criteria can be used to quantify entanglement \cite{Young2006}.

\subsection{Single-photon sources}\label{sec:SPS}
A prominent strategy to increase the photon extraction efficiency from quantum emitters is their integration into microresonators \cite{Barnes2002}.
Cylindrical Fabry-P\'{e}rot microcavitites, also known as micropillar cavities, with embedded stochastically grown QDs have already been exploited in a variety of experiments \cite{Reitzenstein2010} including strong light-matter coupling \cite{Reithmaier2004}, quantum key distribution \cite{Heindel2012}, and boson sampling \cite{Wang2017}. In all these experiments, typically many devices needed to be scanned to find QD-micropillars with sufficient emitter-mode coupling due to the spatially and spectrally random distribution of the emitters inside the cavity. 

Today, deterministic fabrication techniques allow for precisely aligning quantum emitters to micropillar structures, or vice versa, resulting in device yields close to unity with optimized performance. In 2009 Schneider et al. demonstrated the first deterministically fabricated QD-micropillar device using site-controlled QD growth in a bottom-up fashion \cite{Schneider2009} (cf. section~\ref{sec:Bottom-Up}).
\begin{figure}[t]
\begin{center}
\includegraphics[width=1.0\linewidth]{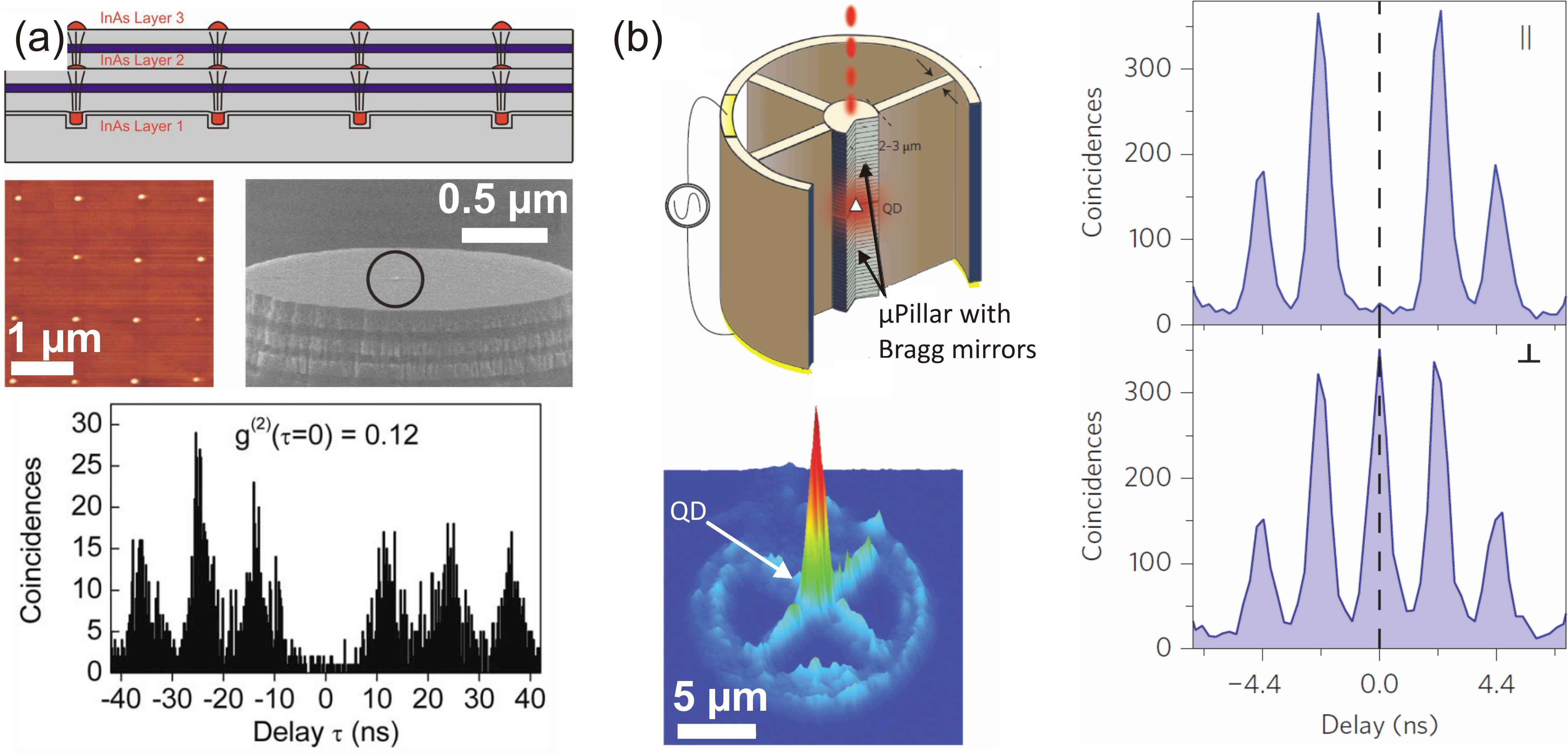}
\caption{Deterministically fabricated single-photon sources based on QD-micropillars using (a) site-controlled QD growth with a spatially aligned device (adapted  from \cite{Schneider2009}, with the permission of AIP Publishing) and (b) in-situ photolithography (\cite{Somaschi2016}, adapted by permission from Springer Nature, \copyright 2016 Macmillan Publishers Limited).}
\label{fig:Fig_6}
\end{center}
\end{figure}
To achieve high optical quality of the site-controlled QDs, two layered arrays of InAs QDs were grown on top of a nanohole seeding-layer each separated by a GaAs buffer layer. Photoluminescence of the seeding-layer as well as of the first QD layer was spectrally blue-shifted by about 30\,nm relative to the third QD layer using partial capping and annealing. The ordered arrays of site-controlled QDs were embedded in a $\lambda$-thick GaAs cavity sandwiched between a lower and an upper distributed Bragg reflector containing 25 and 12 mirror pairs, respectively. Next, micropillars have been fabricated spatially aligned with single positioned QDs using alignment markers. This approach resulted in deterministically fabricated QD-micropillar cavities each containing a single quantum emitter with high yield ($\approx$90\%). The tested device with a pillar diameter of 1\,$\mu$m featured a Q-factor of 1700, which enabled the observation of Purcell-enhanced triggered single-photon emission. The implementation of a top-down approach for the deterministic fabrication of single-photon sources was reported in 2013 by Gazzano et al. \cite{Gazzano2013}. Using the in-situ photolithography developed in Ref. \cite{Dousse2008} (cf. figure \ref{fig:Fig_4}(a)), the authors deterministically fabricated a micropillar cavity containing a single pre-selected QD. This resulted in the observation of large photon extraction efficiencies and high degrees of photon indistinguishability. This technology has been further improved by integrating gates in p-i-n doped micropillars, enabling a spectral tuning of the emission of integrated single quantum emitter relative to the cavity mode \cite{Nowak2014}. These improvements finally culminated in the report on a near-optimal single-photon source by Somaschi et al. \cite{Somaschi2016} and similar results from another group \cite{Unsleber2016}. In the first report, single-photon indistinguishabilities of up to $(99.56\pm0.45)$\% were observed, while the second work achieved extraction efficiencies of up to $(74\pm4)$\%. Besides the deterministic device technology itself, an important key for achieving large indistinguishabilities and photon extraction efficiencies was the resonant excitation scheme \cite{Muller2007} used in both reports. This enables the on-demand generation of single-photon states with near-unity generation probability while keeping dephasing at a minimum.

\begin{figure}[t]
\begin{center}
\includegraphics[width=0.5\linewidth]{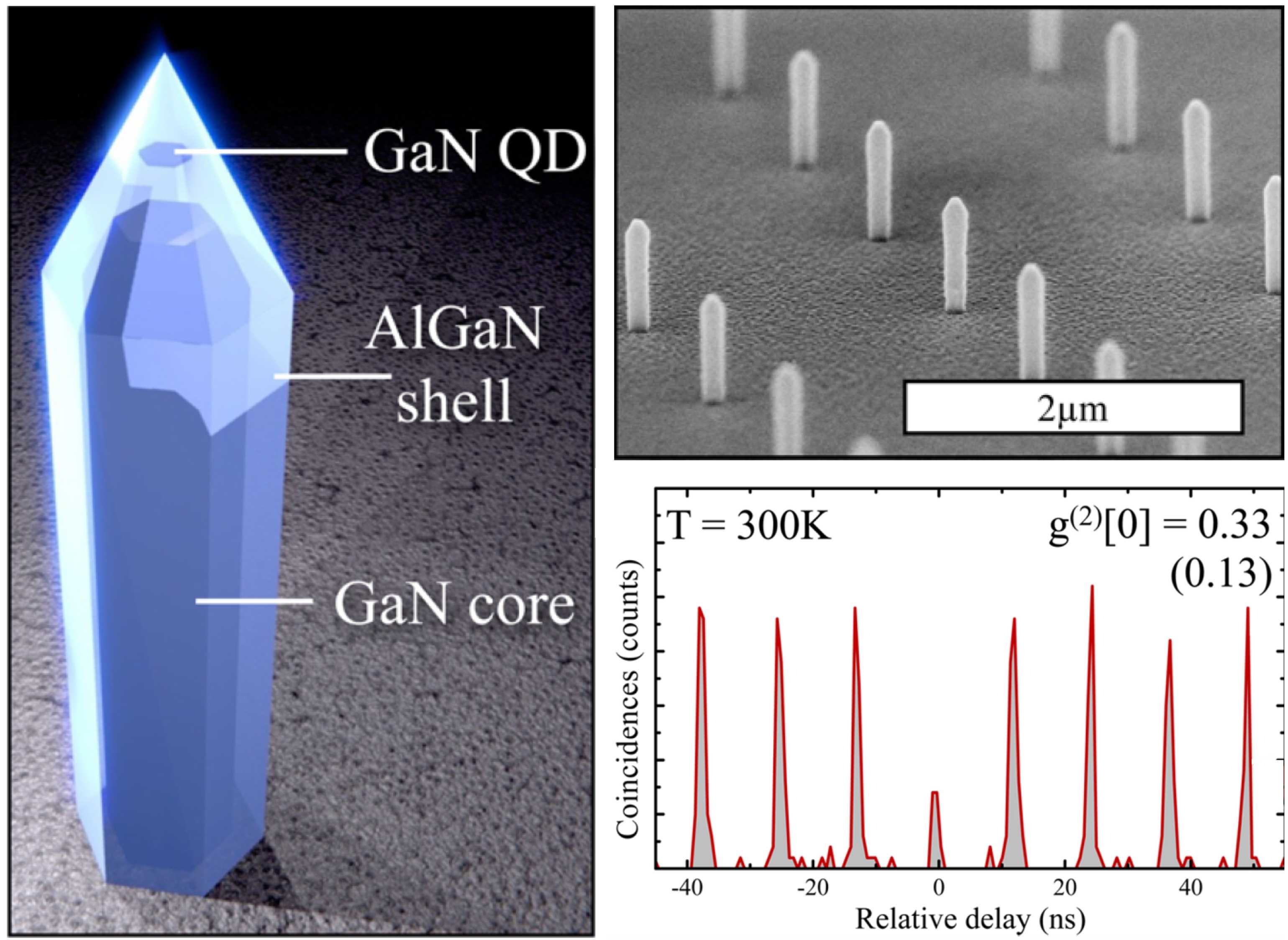}
\caption{Deterministically fabricated nanowire-QD single-photon sources operable up to room-temperature (\cite{Holmes2014}, adapted with permission, \copyright 2014 American Chemical Society).}
\label{fig:Fig_7}
\end{center}
\end{figure}
A drawback of the micropillar strategy discussed above is their narrow bandwidth character. As a consequence the precise spectral matching between cavity mode and quantum emitter often requires an additional tuning-knob, such as temperature, strain or electric field. In order to support high photon extraction efficiencies in a wider spectral window, nanowires \cite{Claudon2010}, lens structures \cite{Hadden2010,Siyushev2010} and circular Bragg gratings \cite{Ates2012a} can been exploited. In 2012 Reimer et al. demonstrated a bright single-photon source based on bottom-up grown tapered InP nanowires with integrated positioned InAsP QDs. Although the spatial distribution of the nanowires  was statistically random in this work, each nanowire contained a single precisely aligned QD. After the growth, the QD-nanowires were embedded in a transparent polymer and removed from the InP substrate to facilitate the evaporation of a backside gold layer acting as mirror. This approach enabled a photon extraction efficiency of 42\% and a measured antibunching value of $g^{(2)}(0)<0.5$ under continuous wave excitation. A spatially fully deterministic nanowire approach was demonstrated by Holmes et al. using site-controlled growth of GaN/AlGaN nanowires on a pre-patterned sapphire substrate \cite{Holmes2014}. In this approach, the nanowires were arranged in ordered arrays and the tip of the nanowire contained a single GaN QD. Due to the strong quantum-confinement possible in this material system, the approach enabled the observation of single-photon emission in the UV-B spectral window up to room-temperature (300\,K) with a measured $g^{(2)}(0)$-value of 0.33. Other recent work on broadband deterministically fabricated QD-based quantum-light sources use hybrid SIL-based devices \cite{Sartison2017} (cf. section~\ref{sec:Hybrid}) or the lensing effect of metallic nanorings for efficient photon extraction \cite{Trojak2017}, also in combination with solid immersion lenses \cite{Trojak2018}.

\begin{figure}[t]
\begin{center}
\includegraphics[width=0.5\linewidth]{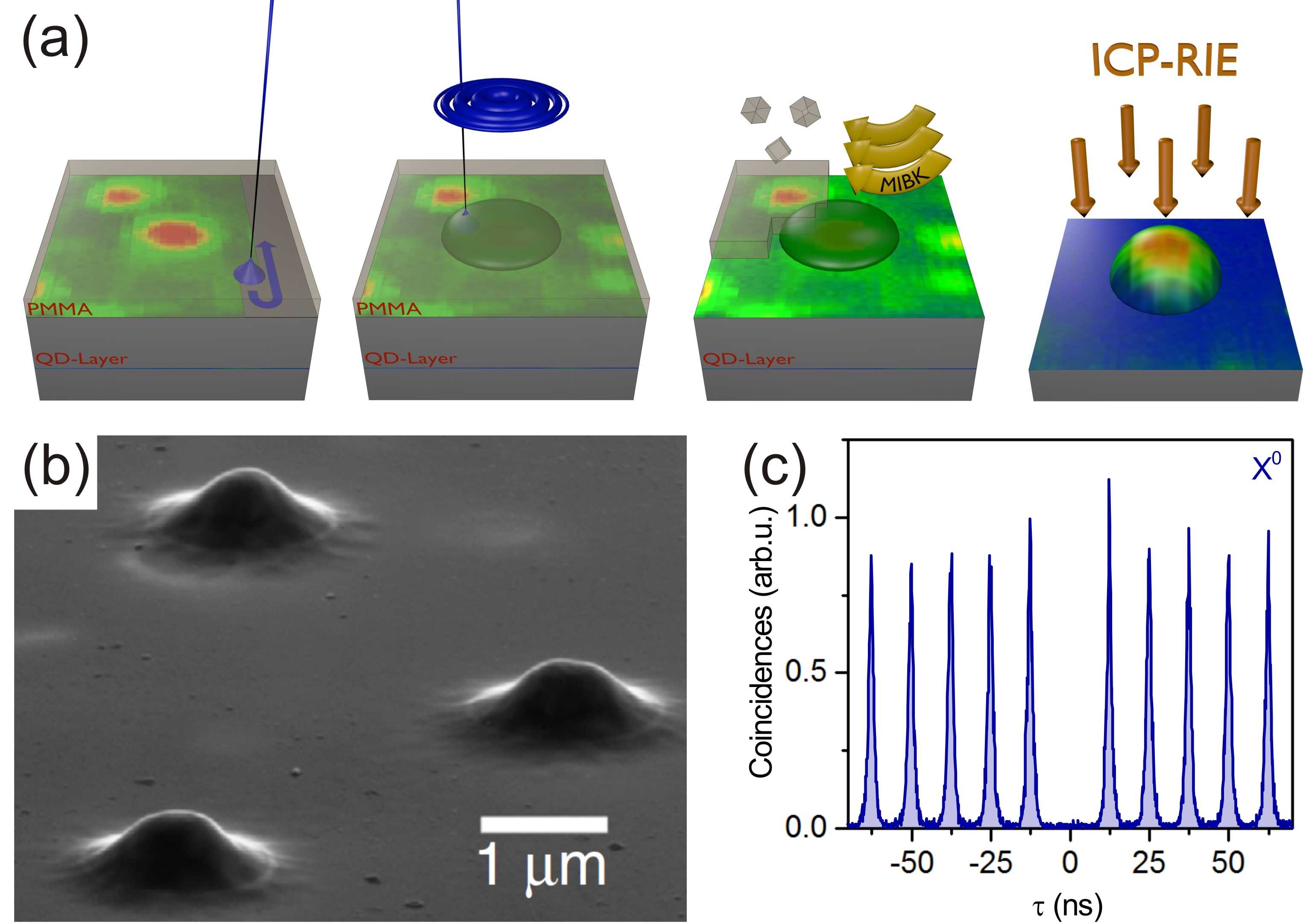}
\caption{Deterministically fabricated single-photon sources based on QD-microlenses using 3D in-situ EBL on pre-selected self-organized grown QDs \cite{Gschrey2015b}.}
\label{fig:Fig_8_1}
\end{center}
\end{figure}

\begin{figure}[t]
\begin{center}
\includegraphics[width=0.5\linewidth]{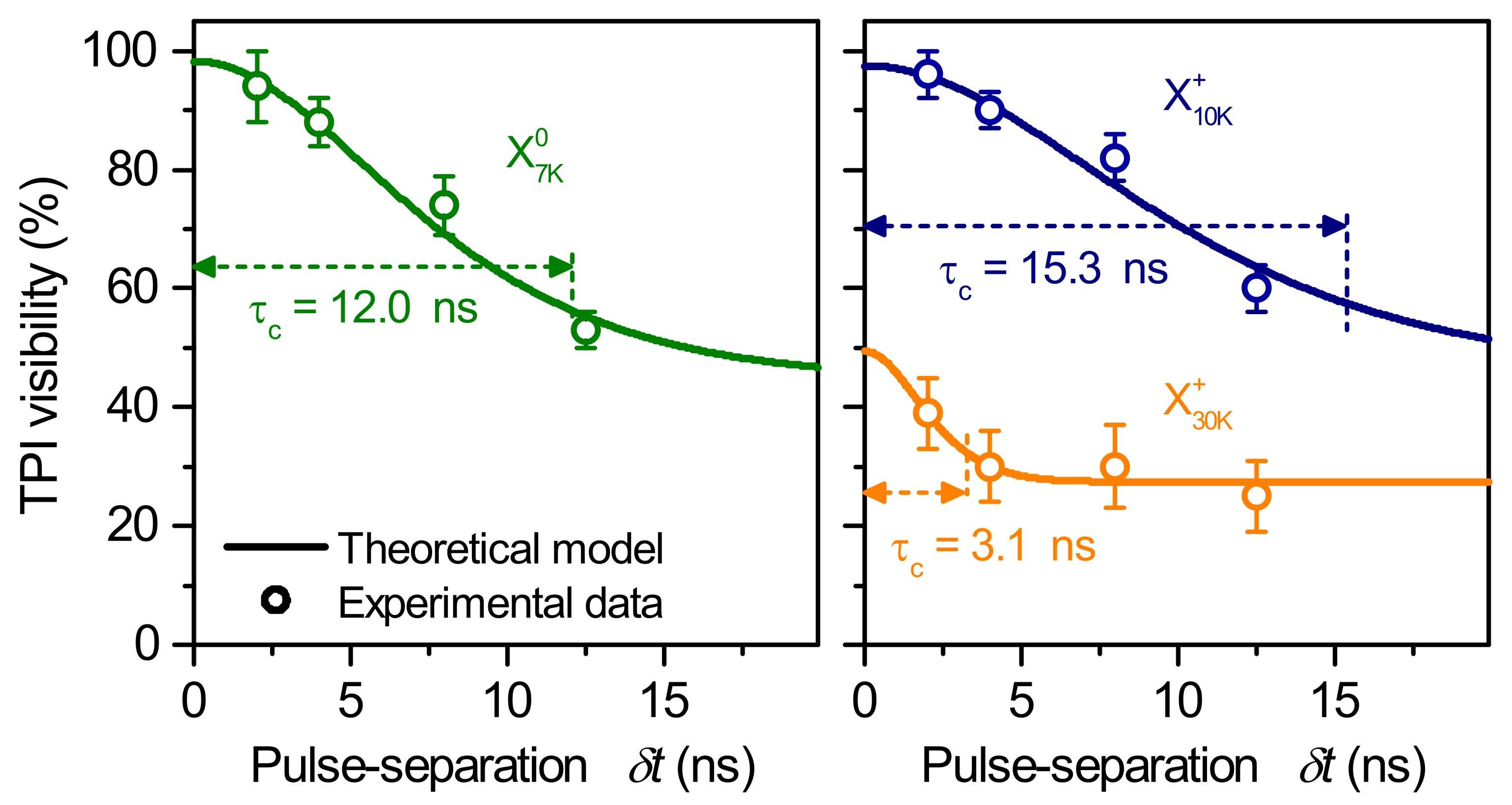}
\caption{Two-photon interference visibilities of consecutively emitted single photons versus the time $\delta t$ elapsed between triggered emission processes. Experimental data for the X$^0$ state (left) and the X$^+$ state (right) are quantitatively described by a theoretical model assuming a non-Markovian noise correlation resulting from spectral diffusion on a nanosecond time scale. A characteristic temperature-dependent correlation time $\tau_C$ is observed. (\cite{Thoma2016}, reprinted figure with permission, \copyright 2016 by the American Physical Society)}
\label{fig:Fig_8_2}
\end{center}
\end{figure}

An particular useful approach was developed in our group by extending the in-situ EBL approach described in section~\ref{sec:In-Situ} to three-dimensional structures. Instead of using a fixed electron dose, resulting in mesa structures as illustrated in figure~\ref{fig:Fig_4}\,(b), we found that a variable dose in the negative-tone regime of the resist (grey-scale lithography) can be used to realize curved surfaces of the photonic structure hosting the single quantum emitter. This has been used by Gschrey et al. to deterministically fabricate a photonic microlens with a single embedded QD acting as bright single photon source \cite{Gschrey2015b,Heindel2017b} (see figure~\ref{fig:Fig_8_1}).
In combination with a back-side DBR the device showed broadband photon extraction efficiencies of $(23\pm3)\%$ into an NA of 0.4, low multi-photon emission probabilities $g^{(2)}(0)<0.01$, and high photon indistinguishabilities, even well beyond saturation of the quantum emitter. In subsequent investigations, we used these QD-microlenses to explore different meachnisms of dephasing limiting the indistinguishability of photons emitted by the quantum emitter \cite{Thoma2016}. In particular we found, that the two-photon interference visibility decreases with increasing temporal separation between consecutively emitted photons (see figure~\ref{fig:Fig_8_2}), which is theoretically described by a non-Markovian noise process. 
At short temporal separations (2\,ns) and low temperatures (10\,K) we observed photon indistinguishabilities of up to $(96\pm4)\%$ under quasi-resonant excitation of the quantum emitter. Furthermore we showed, that the performance of QD-microlens-based SPSs can be further improved in terms of the photon extraction efficieny by employing anti-reflection coatings \cite{Schnauber2016}. Additionally, the achievable single-photon flux has been dramatically increased, by pushing the excitation rate to the limits of the quantum emitters using a mode-locked vertical-external-cavity surface-emitting laser at 500\,MHz repetition rate \cite{Schlehahn2015a}. Beyond the work reported at emission wavelengths of about 900\,nm, also triggered single-photon emission in the telecom O-band has been demonstrated with QD-microlenses \cite{Dusanowski2017}. Meanwhile, the microlens approach has been also adapted by other groups using in-situ photolithography in combination with wet-chemical etching \cite{Sartison2018}.

\subsection{Multi-photon sources}\label{sec:MultiPhoton}
Beyond single-photon generation, the creation of more complex multi-photon states is an extremely exciting and challenging task at the heart of quantum optics. Applications range from quantum repeaters based on sources of entangled photon pairs, quantum-enhanced sensing using N00N-states \cite{Mueller2017} to photonic quantum computing with entangled photonic cluster states \cite{Schwartz2016a}. The increased complexity of the task of distilling a specific multi-photon state, however, typically translates into more demanding boundary conditions for the quantum emitter and its surrounding device. In case of QDs, the XX-X radiative cascade intrinsically offers the possibility to produce highly correlated pairs of photons (cf. section \ref{Sec:QuantumDots}). Due to the typical energy scales for the XX binding energy ($E_{\rm{bin}}^{\rm{XX}}\approx1\,$meV) and the fine-structure splitting of the exciton state ($\Delta E_{\rm{FSS}}\approx10\,\mu$eV), emission of the cascade leads usually to two doublets of orthogonally linearly polarized emission lines visible in the emission spectra, exhibiting spectrally distinguishable photons.

But there exist three particular interesting specific constellations for XX binding energies and fine-structure splittings, respectively. The possibly best known and most studied case corresponds to a QD with a zero fine-structure splitting but a finite XX binding energy. This case results in the emission of polarization-entangled photon pairs \cite{Benson2000,Akopian2006,Stevenson2006}, as discussed in the context of deterministic device fabrication in more detail below. Another type of entanglement which can be produced with the XX-X cascade has been proposed as entanglement via time-reodering, which requires a finite fine-structure splitting but zero XX binding energy \cite{Avron2008}. To the best of our knowledge, this case has not been used for entanglement generation yet. The last case refers to a XX-X cascade, where the fine-structure splitting exactly matches the XX binding energy. While no entanglement is involved here, this specific energy level configuration enables the generation of photon twins - a non-classical light state constituted of two temporally correlated photons with identical emission energy and polarization. In the following we will first discuss experiments based on polarization entanglement using QDs with low fine-structure splittings and then present work on twin-photon emitters.

The first deterministically fabricated device capable of emitting polarization-entangled photon pairs has been demonstrated by Dousse et al. in 2010 in a top-down approach \cite{Dousse2010}. 
\begin{figure}[t]
\begin{center}
\includegraphics[width=0.5\linewidth]{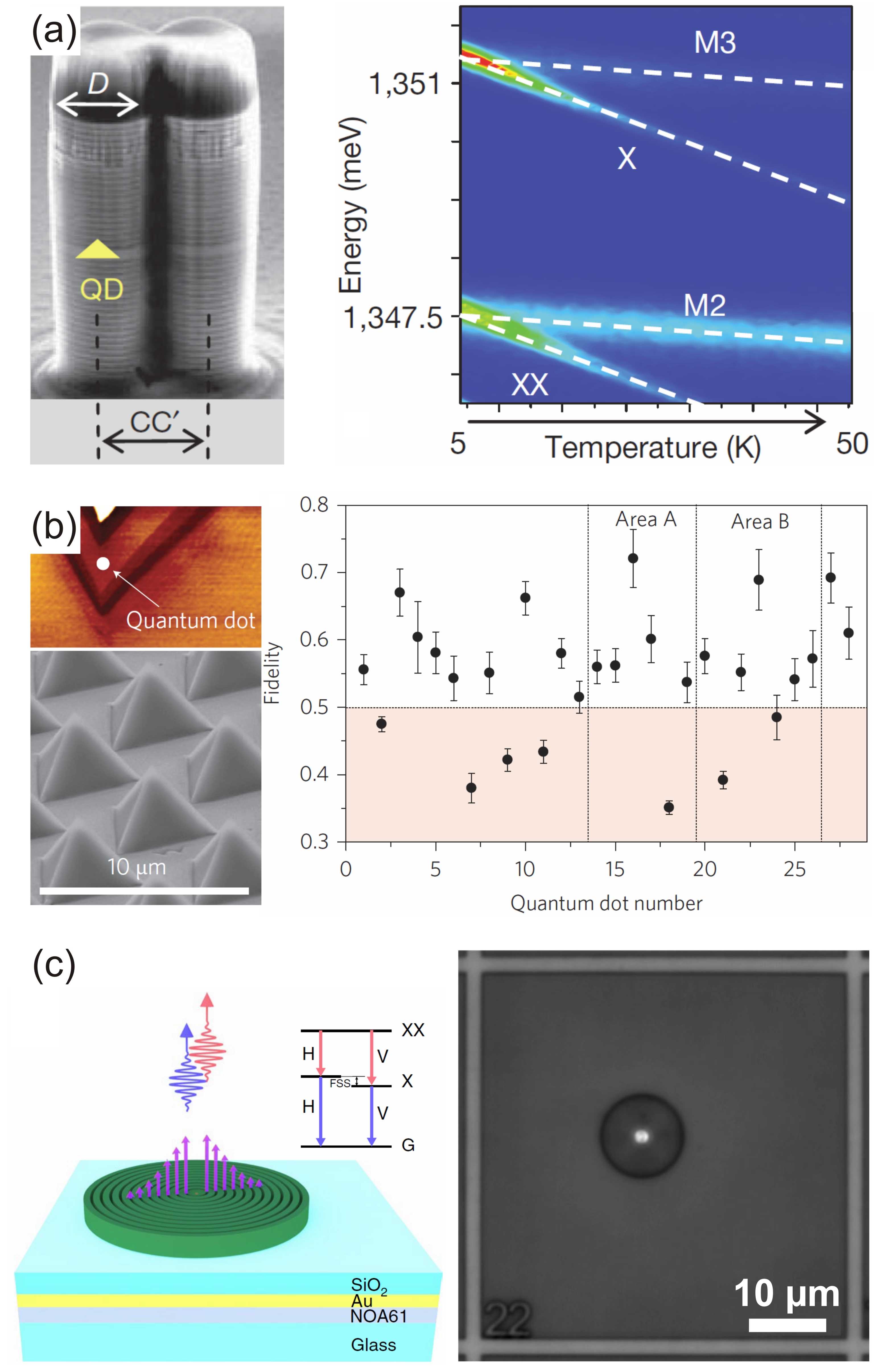}
\caption{Deterministically fabricated QD-based light sources for the generation of polarization-entangled photon pairs: (a) A photonic molecule fabricated around a preselected QD via in-situ photolithography enables efficient photon-pair extraction (\cite{Dousse2010}, reprinted by permission from Springer Nature, \copyright 2010 Macmillan Publishers Limited). (b) Ordered arrays of pyramidal QDs grown on (111)-oriented substrate show average entanglement fidelities significantly above the classical limit (\cite{Juska2013}, reprinted by permission from Springer Nature, \copyright 2013 Macmillan Publishers Limited). (c) A circular Bragg resonator with embedded pre-selected QD in combination with a braodband reflector (\cite{Liu2019}, reprinted by permission from Springer Nature, \copyright 2019 The Authors, under exclusive licence to Springer Nature Limited).}
\label{fig:Fig_Entanglement_1}
\end{center}
\end{figure}
Using in-situ photolithography, a so-called 'photonic molecule' comprising two partially merged micropillars has been processed deterministically around a pre-selected QD (see figure~\ref{fig:Fig_Entanglement_1}(a)). The engineered hybridization of the optical modes of this photonic molecule \cite{Bayer1998} enabled the authors to simultaneously achieve high photon extraction efficiency for exciton and biexciton emission of the integrated QD. This sophisticated approach enabled an entangled photon pair rate of 0.12 per excitation pulse and an entanglement fidelity $F^+$ to the state $\left|\Psi^+\right\rangle$ of 0.59. In a bottom-up approach, Braun et al. observed correlated photon pairs of the XX-X cascade of site-controlled InP/GaInP QDs \cite{Braun2013}. The fine-structure splitting of the fabricated QDs, however, was 300\,$\mu$eV on average, which was too large for entanglement experiments. A successful entanglement experiment based on site-controlled QDs was first reported by Juska et al. by employing pyramidal In$_{0.25}$Ga$_{0.75}$As$_{1-\delta}$N$_{\delta}$ QDs grown on (111)-oriented substrate \cite{Juska2013}. Due to the high symmetry of the realized QDs (see figure~\ref{fig:Fig_Entanglement_1}(b)), fine-structure splittings below 4\,$\mu$eV have been observed for ordered arrays of quantum emitters, as reported also earlier by another group \cite{Mohan2010,Mohan2012}. Entanglement fidelities of up to $0.721\pm0.043$ were observed, while the overall yield of positioned QDs showing $F^+>0.5$ reached 15\%. 

Many of the reports discussed above achieved high entanglement fidelities, but did not satisfy other requirements relevant for entanglement-based quantum information processing. In fact, it is extremely challenging to simultaneously achieve high photon extraction efficiency, high photon indistinguishability and hight entanglement fidelity in a single device. Recently this task has been mastered by Liu et al. using a deterministically fabricated solid-state-based entangled photon pair source comprising a single QD embedded in a circular Bragg resonator integrated on a broadband reflector \cite{Liu2019}. The device (see figure~\ref{fig:Fig_Entanglement_1}(c)), exploiting a broadband Purcell effect, enabled the generation of entangled photon pairs with a collection efficiency of 0.65(4), an entanglement fidelity of 0.88(2), and an indistinguishability of 0.901(3) and 0.903(3) for X and XX respectively. Here, the authors used two-photon resonant excitation \cite{Mueller2014} to achieve the on-demand generation of XX-X photon pairs. Similar results were also reported for a non-deterministically fabricated device by Wang et al. \cite{Wang2019}. Further improvements in the device performance may be possible by using polarization-selective Purcell microcavities \cite{Wang2019a}.

In cases where the fine-structure splitting is large compared to the homogenous linewidth, the polarization entanglement between XX-X photon pairs will be washed out in time-integrated experiments \cite{Stevenson2008}. Still the entanglement can be restored by temporal postselection, if the temporal resolution of the setup is sufficient. Based on this idea, Huber et al. observed polarization-entangled photon pairs with a fidelity of 0.76 emitted by the XX-X cascade of a single QD despite a fine-structure splitting of 18\,$\mu$eV \cite{Huber2014}. The single InAsP QDs were embedded in ordered arrays of tapered InP nanowires. 
To resolve the temporal evolution of the entangled two-photon state, avalanche photo diodes with a temporal resolution of 35\,ps were employed.
While this work was conducted under non-resonant excitation into the conduction band of InP, resonant excitation schemes provide the clear benefit of reduced decoherence.
Employing resonant two-photon excitation of the XX-X cascade of a QD embedded in a deterministically fabricated microlens, Bounouar et al. reported the generation of a maximally entangled state \cite{Bounouar2018}. Here, the oscillating temporal evolution of the polarization-entangled photon pair state was observed for two different quantum emitters with a fine-structure splitting of 16\,$\mu$eV and 30\,$\mu$eV, respectively. Similiar results have been obtained also for non-deterministic device approaches \cite{Winik2017}.

\begin{figure}[t]
\begin{center}
\includegraphics[width=1.0\linewidth]{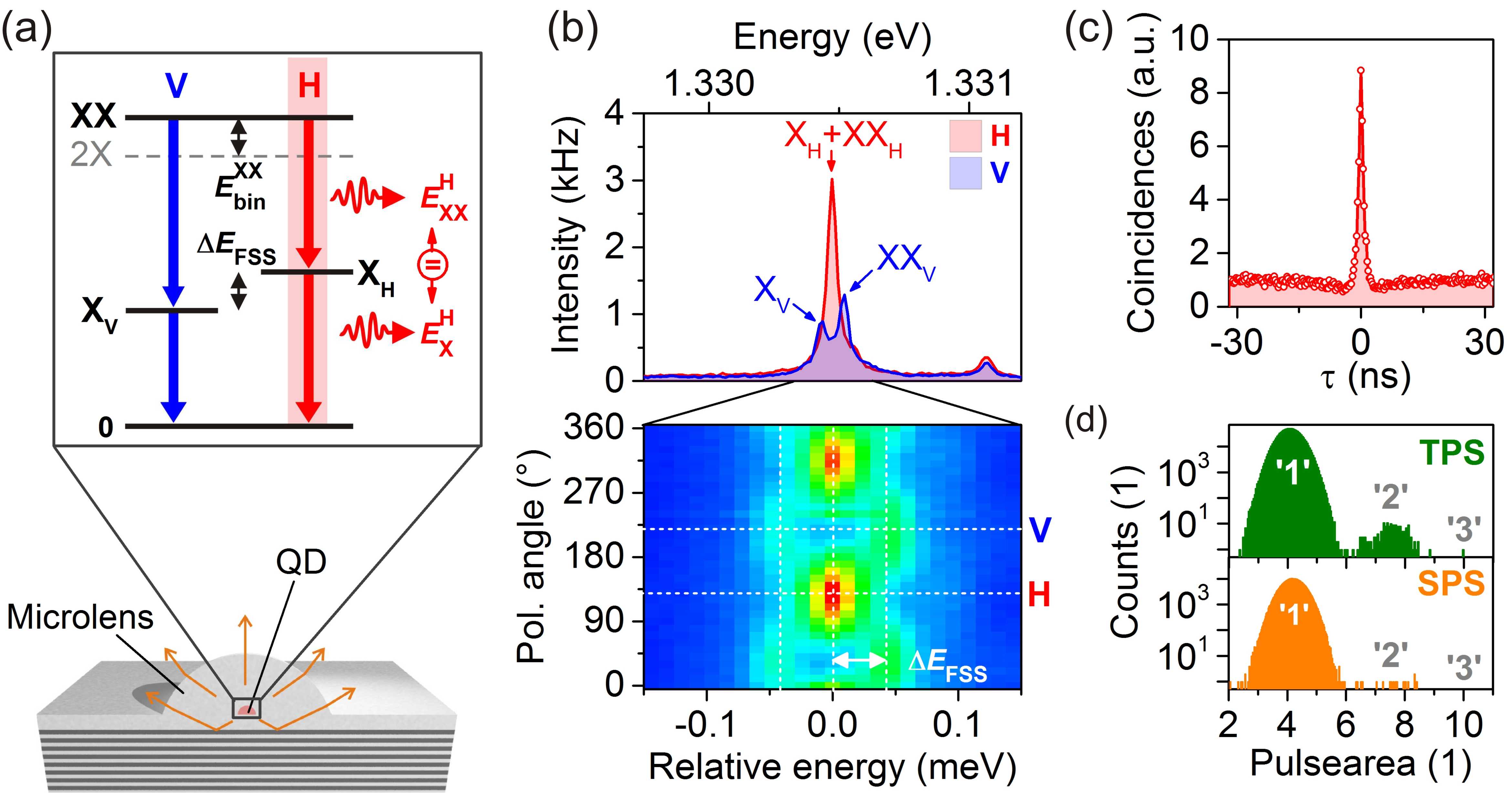}
\caption{Twin-photon generation using deterministically fabricated QD microlenses: (a) Energy level alignment of a quantum dot exhibiting a XX-X radiative cascade with $E_{\rm{FSS}}=\left|E_{\rm{bin}}^{\rm{XX}}\right|$. The photon pairs emitted in the H-polarized decay path have identical energy and polarization. (b) Polarization-resolved $\mu$PL spectra of a twin-photon cascade as sketched in (a). (c) Photon-autocorrelation measurement on the H-polarized decay channel. (d) Photon number distribution of the twin-photon source (TPS) and a single-photon source (SPS), serving as reference, deduced from measurements using photon-number-resolving detectors. (adapted from \cite{Heindel2017}, CC BY 4.0)}
\label{fig:Fig_TwinPhoton}
\end{center}
\end{figure}

In the experiments discussed above, the biexciton binding energy was large compared to the fine-structure splitting. This energy level alignment present in most QDs leads to spectrally separated emission lines from the XX- and X-state, respectively. Recently we demonstrated for the first time, that also the special case of an energetically degenerate XX-X cascade can be realized with QDs \cite{Heindel2017}. For this purpose, we selected deterministically fabricated QD microlenses with a XX-X cascade satisfying the condition $\Delta E\rm{_{FSS}}$=$|E\rm{_{bin}^{XX}}|$. For this specific case, the X- and the XX-photon have the same emission energy for either H or V polarization, while they are spectrally separated by twice the fine-structure splitting for the respective orthogonal polarization (see figure~\ref{fig:Fig_TwinPhoton}\,(a)). 
Polarization filtering of the energetically degenerate decay path thus enables the distillation of photon twins - a non-classical light state constituted of two temporally correlated photons with identical emission energy and polarization. Figure~\ref{fig:Fig_TwinPhoton}\,(b) presents polarization dependent $\mu$PL spectra of a QD, showing such a behavior.
The high degree of temporal correlations present in the emitted twin-photon state leads to a huge bunching effect in photon auto-correlation measurements using a HBT setup (see figure~\ref{fig:Fig_TwinPhoton}\,(c)).
A more direct observation of the photon twins was possible in this work, by employing a photon-number-resolving detection system based on superconducting transition edge sensors. As displayed in Figure~\ref{fig:Fig_TwinPhoton}\,(d), we were able to clearly identify the events originating from the emission of photon twins and, further more, directly compare the photon number distribution to a QD-based single-photon source serving as reference. In a very recent study, Moroni et al. picked up this idea and studied similiar cases in site-controlled pyramidal QDs \cite{Moroni2019}.

Going one step further to generate correlated three-photon states, Khoshnegar et al. used vertically-stacked coupled QDs integrated in site-controlled nanowires \cite{Khoshnegar2017}. In this approach, the hybridized excitonic states of the two QDs form a triexciton-biexciton-exciton cascade, consecutively emitting three photons with different energies. Noteworthy, the generation of three-photon states can also be achieved in a single QD using its triexciton-biexciton-exciton radiative cascade, as demonstrated by Schmidgall et al. in a non-deterministic device approach \cite{Schmidgall2014a}. And even four-photon states can be generated using a single QD emitter in this fashion \cite{Arashida2011}.

\subsection{Electrically-driven quantum-light sources}\label{sec:EL}
\begin{figure}[t]
\begin{center}
\includegraphics[width=1.0\linewidth]{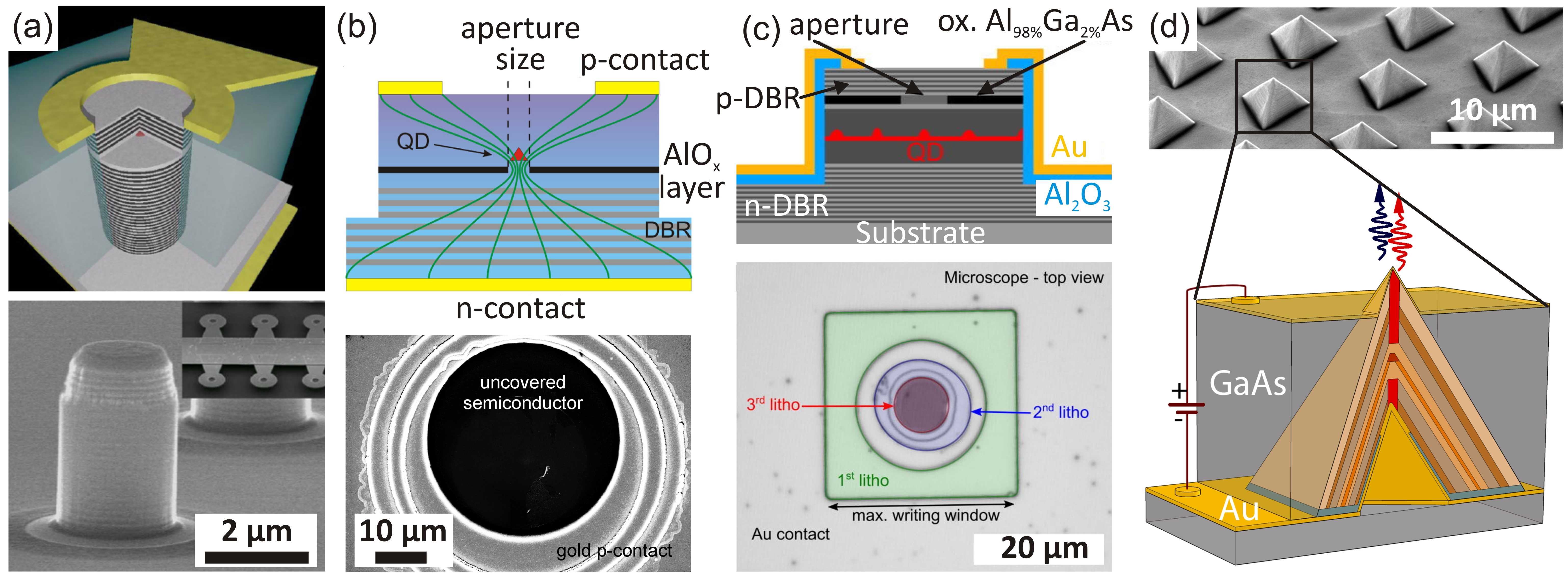}
\caption{Electrically-driven deterministically-fabricated QD-based quantum-light sources: (a) Micropillar cavity with an integrated site-controlled QD (adapted from \cite{Schneider2012}, with the permission of AIP Publishing), (b) QD light emitting diode with a positioned QD based on a buried oxide aperture (adapted from \cite{Unrau2012}, with the permission of AIP Publishing), (c) QD light emitting diode based on in-situ photolithography of pre-selected stochastically grown QDs (adapted from \cite{Sartison2019}, with the permission of AIP Publishing)), and (d) Entangled-photon light emitting diode based on ordered arrays of pyramidal QDs (\cite{Chung2016}, adapted by permission from Springer Nature, \copyright 2016 Macmillan Publishers Limited).}
\label{fig:Fig_9}
\end{center}
\end{figure}

Early work on deterministically-fabricated electrically-driven single-QD light emitting diodes confirmed the prospects for device integration \cite{Baier2004,Mehta2010,Huggenberger2011a}, a proof that the electrically-pumped QD emission was antibunched, however, was still missing. The first electrically-driven quantum-light source based on a deterministic fabrication technology was reportet in 2012 by Schneider et al. \cite{Schneider2012}. In this work, single site-controlled InAs QDs have been integrated in micropillar cavities incorporating a p-i-n doped diode structure (see Figure~\ref{fig:Fig_9}\,(a)).
The chosen sample layout comprising 25 and 5 AlAs/GaAs mirror pairs in the lower and upper DBR, respectively, enabled a Q-factor of 230. This enabled the authors to demonstrate single-photon emission with $g^{(2)}(0)=0.42$ (considering the temporal resolution of the experimental setup) of the Purcell-enhanced emission of a deterministically integrated QD under direct current injection. Due to the nanohole seeding layer in close vicinity to the site-controlled QDs, the observed linewidth of the QD emission was about one order of magnitude larger as compared to high-quality stochastically grown QDs in this material system. Unrau et al. demonstrated in Ref. \cite{Unrau2012} a QD single-photon emitting diode with significantly improved optical properties based on site-controlled QDs positioned via a buried oxide aperture (cf. figure~\ref{fig:Fig_2}\,(c) in section~\ref{sec:Bottom-Up}). The device illustrated in figure~\ref{fig:Fig_9}\,(b) showed resolution limited linewidths of 25\,$\mu$eV and pronounced antibunching of $g^{(2)}(0)=0.05$.
More recently, a deterministically fabricated single-photon light emitting diode has also been realized based on the pre-selection of self-organized QDs and subsequent device fabrication using in-situ photolithography \cite{Sartison2019} (see figure~\ref{fig:Fig_9}\,(c)). The authors achieved a measured antibunching of $g^{(2)}(0)=0.42\pm0.02$ under pulsed electrical current injection at 200\,MHz.
The first deterministically-fabricated entangled-light-emitting diode was reported by Chung et al. based on arrays of pyramidal QDs grown on (111)B oriented GaAs substrate \cite{Chung2016} (see figure~\ref{fig:Fig_9}\,(d)). As discussed already in section~\ref{sec:Bottom-Up}, this type of quantum emitter intrinsically shows small fine-structure splitting for the bright exciton state and is thus well suited for the generation of polarization-entangled photo pairs. The device showed entanglement fidelities of the emitted two-photon state of $0.73\pm0.06$ under direct current injection and $0.678\pm0.023$ for pulsed current injection at 63\,MHz.

\subsection{Plug-and-play quantum-light sources}\label{sec:PnP}
\begin{figure}[t]
\begin{center}
\includegraphics[width=1.0\linewidth]{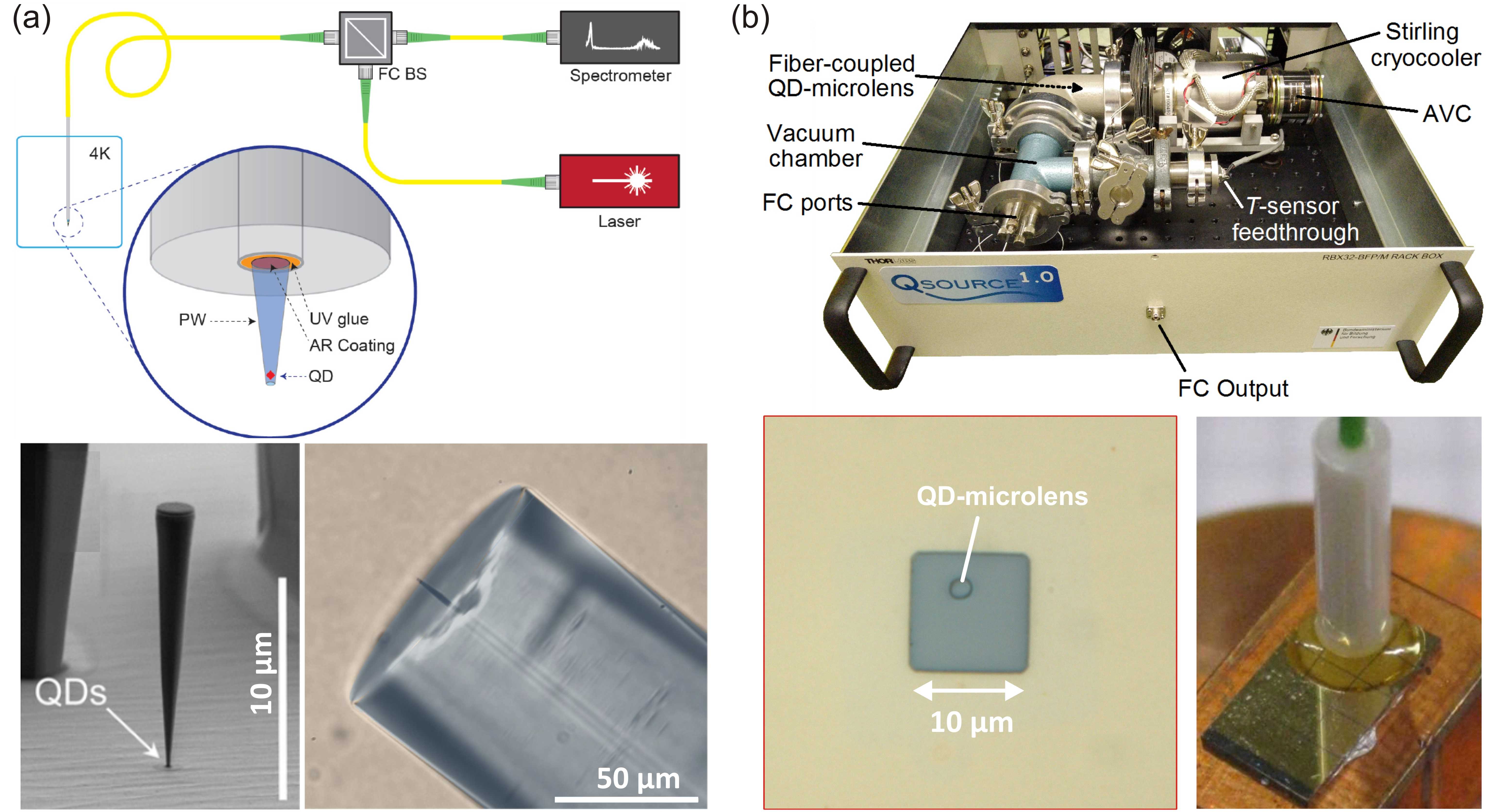}
\caption{Development of plug-and-play single-photon sources with direct fiber coupling based on (a) a nanowire QD attached to the fiber core via a pick-and-place technique (adapted from \cite{Cadeddu2016}, with the permission of AIP Publishing) and (b) a fiber-coupled deterministically fabricated QD microlens integrated in a compact Stirling cryocooler (\cite{Schlehahn2018}, CC BY 4.0).}
\label{fig:Fig_10}
\end{center}
\end{figure}

In view of the huge progress achieved in the fabrication of deterministic devices at the chip level (see previous sections), recent efforts aim also at the development of more practical plug-and-play devices. To achieve this goal, a direct fiber coupling of the respective quantum devices is beneficial. Pioneering work in this direction has been performed by Xu et al. in Ref.~\cite{Xu2008} and Haupt et al. in Ref.~\cite{Haupt2010} using non-deterministic device approaches. More recently, Caddedu et al. reported in Ref. \cite{Cadeddu2016} a fiber-coupled QD on a photonic tip (see figure~\ref{fig:Fig_10}\,(a)) using deterministic technologies. 
For this purpose, a so-called photonic trumpet \cite{Munsch2013} with an embedded InAs QD was precisely attached to the core of a single mode optical fiber using a pick-and-place method with custom-made micro-manipulators inside a SEM. For testing single-photon emission, the fiber-coupled quantum emitter was plunged into liquid helium and connected to a fiber-based PL setup. Under continuous wave (CW) wetting-layer excitation, a clear antibunching $g^{(2)}(0)<0.5$ was observed in a HBT experiment, which was only limited by the setup's temporal resolution. To develop quantum-light sources which are even more practical, one has either to get rid of the necessity for cooling, by working on the material's side, or one applies more practical techniques for cooling the quantum emitter. Although room-temperature operation of QD single-photon sources has been demonstrated (see section \ref{sec:SPS} and Ref. \cite{Holmes2014}), the quantum-optical properties of the generated photons and the achieved brightness is so far insufficient for most applications in photonic quantum technologies. Stirling cryocoolers \cite{Veprik2005} have been evaluated as compact, cost-effective, and user-friendly alternative to liquid-helium bath or flow cryostats and closed-cycle refrigerators. The applicability of Stirling cryocoolers for the operation of quantum emitters has first been demonstrated in 2015 by Schlehahn et al., using a deterministically fabricated QD mesa cooled to a base-temperature of about 29\,K \cite{Schlehahn2015}. Based on this idea, we recently presented a stand-alone fiber-coupled single-photon source \cite{Schlehahn2018}, comprising a deterministically-fabricated fiber-coupled QD single-photon source inside a compact Stirling cryocooler fitting into a 19-inch rack box (see figure~\ref{fig:Fig_10}\,(b)). The fiber-coupled QD emission provides the user with a photonflux stable on a time-scale of several days and antibunching values as small as $g^{(2)}(0)=0.07\pm0.05$ under CW optical excitation (using external spectral filtering). Additionally the high durability of the direct fiber-connection was demonstrated in endurance tests, revealing stable collection of the QD emission within 4\% over several cool-down/warm-up cycles. Our device thus demonstrated for the first time, that practical high-performance plug-and-play quantum-light sources suitable for applications outside shielded lab environments are within reach. The efficiency of this first demonstrator in terms of the photon extraction probability per pulse, however, was below 1\%, limited by the fiber-coupling efficiency and the photon extraction efficiency of the used microlens device. A crucial next step in this direction will therefore be to improve the coupling efficiency to single-mode optical fibers in such modular and robust approaches, e.g. by deterministically integrating micropillar cavities \cite{Snijders2018} or numerically optimized circular Bragg gratings for operation at telecom wavelengths \cite{Rickert2019}. 
Another important step will be the implementation of coherent excitation schemes for portable quantum-light sources. Many reports on state-of-the-art quantum-light sources use strict-resonant \cite{Somaschi2016,Unsleber2016,Wang2019a}, two-photon resonant \cite{Bounouar2018,Liu2019,Wang2019}, or, more recently, dichromatic coherent excitation \cite{He2019} for the on-demand generation of quantum light states (cf. section~\ref{sec:QuantumLightSources}). The implementation of such schemes in compact and integrated approaches, however remains challenging, as the required laser systems and pulse-shapers are rather bulky. In this regard the development of compact and tunable picosecond pulsed laser systems would be an important technological achievement, further boosting the realization of practical quantum-light sources.

\section{Applications and future perspectives}\label{sec:Applications}
The progress achieved in the development of deterministic fabrication techniques has significantly improved the device yield and the performance of quantum-light sources. During the next decade, this development will push applications of photonic quantum technologies far beyond its current status. To date, most proof-of-concept experiments in the field of quantum communication and quantum computing have been demonstrated based on non-deterministic device approaches. In case of QDs, examples include quantum key distribution (QKD) via the BB84 protocol \cite{Waks2002,Intallura2007,Takemoto2010,Collins2010,Heindel2012}, spin-photon\cite{Gao2012,DeGreve2012}, spin-spin \cite{Delteil2015} and photon-photon \cite{Schwartz2016a} entanglement as well as boson sampling \cite{Wang2017}. Also NV and SiV colour centres in diamond crystals have been used for QKD proof-of-principles \cite{Beveratos2002,Alleaume2004,Leifgen2014} and tests of Bell inequality \cite{Hensen2015}. 

To surpass the proof-of-principle stage and to realize systems with increased complexity, however, scalable device technologies are crucial. Building multi-node quantum communication networks or photonic quantum computers, for instance, many indistinguishable quantum-light sources are required. Therefore, these applications might well be the first, where deterministically fabricated single-photon and entangled photon-pair sources make the difference. First experiments in photonic quantum technologies exploiting deterministic devices include a boson sampling machine using micropillar single-photon sources \cite{Loredo2017}. Moreover, two-photon interference experiments on remote deterministically fabricated micropillar single-photon sources \cite{Giesz2015} as well as QD microlenses \cite{Thoma2016} were demonstrated, which rise prospects for scalable implementations of device-independent schemes of QKD \cite{Lo2012,Lo2014} and quantum repeater networks.
Furthermore, also the development of devices integrated \textit{on-chip} will strongly benefit from deterministic device technologies. Building on earlier non-deterministic device approaches \cite{Prtljaga2014, Rengstl2015}, we recently demonstrated an on-chip HBT experiment using a single pre-selected QD deterministically integrated into a monolithic waveguide circuit comprising a multi-mode interference beamsplitter \cite{Schnauber2018}. The substantially improved device yield of such deterministic approaches raises prospects for the fabrication of complex photonic circuits on-chip. Not least, also the field of quantum metrology can be boosted by deterministic devices. Here, the development of absolute single-photon sources and the definition of the SI unit Candela will be particularly benefitting by the high degree of scalability of quantum-light sources.

\section{Acknowledgments}
We gratefully acknowledge financial support from the German Federal Ministry of Education and Research (BMBF) via the project \textit{QuSecure} (Grant No. 13N14876) within the funding program Photonic Research Germany and the German Research Foundation (DFG) via the SFB 787 project \textit{Semiconductor Nanophotonics: Materials, Models, Devices}.

\clearpage

\bibliography{Literatur}

\providecommand{\newblock}{}
\begin{thebibliography}{100}
\expandafter\ifx\csname url\endcsname\relax
  \def\url#1{{\tt #1}}\fi
\expandafter\ifx\csname urlprefix\endcsname\relax\def\urlprefix{URL }\fi
\providecommand{\eprint}[2][]{\url{#2}}

\bibitem{OBrien2009}
O'Brien J~L, Furusawa A and Vuckovic J 2009 {\em Nat. Photon.\/} {\bf 3}(12)

\bibitem{Gisin2002}
Gisin N, Ribordy G, Tittel W and Zbinden H 2002 {\em Rev. Mod. Phys.\/} {\bf
  74} 145--195

\bibitem{Gisin2007}
Gisin N and Thew R 2007 {\em Nat. Photon.\/} {\bf 1} 165--171

\bibitem{Lo2014}
Lo H~K, Curty M and Tamaki K 2014 {\em Nat. Photon.\/} {\bf 8} 595--604

\bibitem{Flamini2018}
Flamini F, Spagnolo N and Sciarrino F 2018 {\em Reports on Progress in
  Physics\/} {\bf 82} 016001

\bibitem{Bennett1984}
Bennett C~H and Brassard G 1984 {\em Proceedings of IEEE International
  Conference on Computers, Systems and Signal Processing, Bangalore, India\/}
  175--179

\bibitem{Briegel1998}
Briegel H~J, D\"ur W, Cirac J~I and Zoller P 1998 {\em Phys. Rev. Lett.\/} {\bf
  81}(26) 5932--5935

\bibitem{Sangouard2007}
Sangouard N, Simon C, Min\'a\ifmmode~\check{r}\else \v{r}\fi{} J, Zbinden H,
  de~Riedmatten H and Gisin N 2007 {\em Phys. Rev. A\/} {\bf 76}(5) 050301

\bibitem{Keizer2012}
Keizer J~G, Henriques A~B, Maia A~D~B, Quivy A~A and Koenraad P~M 2012 {\em
  Appl. Phys. Lett.\/} {\bf 101} 243113

\bibitem{Jelezko2006}
Jelezko F and Wrachtrup J 2006 {\em Phys. Status Solidi A\/} {\bf 203}
  3207--3225

\bibitem{Ma2015a}
Ma X, Hartmann N~F, Baldwin J~K~S, Doorn S~K and Htoon H 2015 {\em Nat.
  Nanotechnol.\/} {\bf 10} 671--675

\bibitem{Radisavljevic2011}
Radisavljevic B, Radenovic A, Brivio J, Giacometti V and Kis A 2011 {\em Nat.
  Nanotechnol.\/} {\bf 6} 147--150

\bibitem{Tonndorf2015}
Tonndorf P, Schmidt R, Schneider R, Kern J, Buscema M, Steele G~A,
  Castellanos-Gomez A, van~der Zant H~S~J, Michaelis~de Vasconcellos S and
  Bratschitsch R 2015 {\em Optica\/} {\bf 2} 347

\bibitem{Leonard1993}
Leonard D, Krishnamurthy M, Reaves C~M, Denbaars S~P and Petroff P~M 1993 {\em
  Appl. Phys. Lett.\/} {\bf 63} 3203--3205

\bibitem{Grundmann1995}
Grundmann M, Christen J, Ledentsov N~N, B\"ohrer J, Bimberg D, S R~S, Werner P,
  Richter U, G\"osele U, Heydenreich J, Ustinov V~M, Egorov A~Y, Zhukov A~E,
  Kop\'ev P~S and Alferov Z~I 1995 {\em Phys. Rev. Lett.\/} {\bf 74} 4043--4046

\bibitem{Arakawa1982}
Arakawa Y and Sakaki H 1982 {\em Appl. Phys. Lett.\/} {\bf 40} 939--941

\bibitem{Imamoglu1999}
Imamoglu A, Awschalom D~D, Burkard G, DiVincenzo D~P, Loss D, Sherwin M and
  Small A 1999 {\em Phys. Rev. Lett.\/} {\bf 83} 4204--4207

\bibitem{Michler2009}
Michler P 2009 {\em Single Semiconductor Quantum Dots\/} (Springer)

\bibitem{Lay1978}
Lay G~L and Kern R 1978 {\em J. Cryst. Growth.\/} {\bf 44} 197--222

\bibitem{Petroff2001}
Petroff P~M, Lorke A and Imamoglu A 2001 {\em Physics Today\/} {\bf 54} 46--52

\bibitem{Pohl2013}
Pohl U 2009 {\em Epitaxy of Semiconductors\/} (Springer-Verlag Berlin
  Heidelberg)

\bibitem{Borri2001}
Borri P, Langbein W, Schneider S, Woggon U, Sellin R~L, Ouyang D and Bimberg D
  2001 {\em Phys. Rev. Lett.\/} {\bf 87}(15) 157401

\bibitem{Besombes2001}
Besombes L, Kheng K, Marsal L and Mariette H 2001 {\em Phys. Rev. B\/} {\bf 63}

\bibitem{Bayer2002}
Bayer M and Forchel A 2002 {\em Phys. Rev. B\/} {\bf 65}(4) 041308

\bibitem{Gammon1996}
Gammon D, Snow E~S, Shanabrook B~V, Katzer D~S and Park D 1996 {\em Phys. Rev.
  Lett.\/} {\bf 76}(16) 3005--3008

\bibitem{Kulakovskii1999}
Kulakovskii V~D, Bacher G, Weigand R, K\"{u}mmell T, Forchel A, Borovitskaya E,
  Leonardi K and Hommel D 1999 {\em Phys. Rev. Letters\/} {\bf 82} 1780--1783

\bibitem{Rodt2003}
Rodt S, Heitz R, Schliwa A, Sellin R~L, Guffarth F and Bimberg D 2003 {\em
  Phys. Rev. B\/} {\bf 68}(3) 035331

\bibitem{Sarkar2006}
Sarkar D, van~der Meulen H~P, Calleja J~M, Becker J~M, Haug R~J and Pierz K
  2006 {\em J. Appl. Phys.\/} {\bf 100} 023109

\bibitem{Seguin2006}
Seguin R, Schliwa A, Germann T~D, Rodt S, Pötschke K, Strittmatter A, Pohl
  U~W, Bimberg D, Winkelnkemper M, Hammerschmidt T and Kratzer P 2006 {\em
  Appl. Phys. Lett.\/} {\bf 89} 263109

\bibitem{Brown1956}
Hanbury-Brown R and Twiss R~Q 1956 {\em Nature\/} {\bf 177} 27 -- 29

\bibitem{Schweickert2018}
Schweickert L, J\"{o}ns K~D, Zeuner K~D, Covre~da Silva S~F, Huang H, Lettner
  T, Reindl M, Zichi J, Trotta R, Rastelli A and Zwiller V 2018 {\em Appl.
  Phys. Lett.\/} {\bf 112} 093106

\bibitem{Regelman2001}
Regelman D~V, Mizrahi U, Gershoni D, Ehrenfreund E, Schoenfeld W~V and Petroff
  P~M 2001 {\em Phys. Rev. Lett.\/} {\bf 87}(25) 257401

\bibitem{Kiraz2002}
Kiraz A, F\"alth S, Becher C, Gayral B, Schoenfeld W~V, Petroff P~M, Zhang L,
  Hu E and Imamo\u{g}lu A 2002 {\em Phys. Rev. B\/} {\bf 65} 161303

\bibitem{Santori2002a}
Santori C, Fattal D, Pelton M, Solomon G~S and Yamamoto Y 2002 {\em Phys. Rev.
  B\/} {\bf 66}(4) 045308

\bibitem{Benson2000}
Benson O, Santori C, Pelton M and Yamamoto Y 2000 {\em Phys. Rev. Lett.\/} {\bf
  84} 2513--2516

\bibitem{Schliwa2009a}
Schliwa A, Winkelnkemper M and Bimberg D 2009 {\em Phys. Rev. B\/} {\bf 79}
  075443

\bibitem{Akopian2006}
Akopian N, Lindner N~H, Poem E, Berlatzky Y, Avron J, Gershoni D, Gerardot B~D
  and Petroff P~M 2006 {\em Phys. Rev. Letters\/} {\bf 96} 130501

\bibitem{Young2006}
Young R~J, Stevenson R~M, Atkinson P, Cooper K, Ritchie D~A and Shields A~J
  2006 {\em New J. Phys.\/} {\bf 8} 29

\bibitem{Heindel2017}
Heindel T, Thoma A, von Helversen M, Schmidt M, Schlehahn A, Gschrey M,
  Schnauber P, Schulze J~H, Strittmatter A, Beyer J, Rodt S, Carmele A, Knorr A
  and Reitzenstein S 2017 {\em Nat. Commun.\/} {\bf 8} 14870

\bibitem{Moroni2019}
Moroni S, Varo S, Juska G, Chung T, Gocalinska A and Pelucchi E 2019 {\em J.
  Cryst. Growth.\/} {\bf 506} 36--39

\bibitem{Yuan2002}
Yuan Z, Kardynal B~E, Stevenson R~M, Shields A~J, Lobo C~J, Cooper K, Beattie
  N~S, Ritchie D~A and Pepper M 2002 {\em Science\/} {\bf 295} 102--105

\bibitem{Salter2010}
Salter C~L, Stevenson R~M, Farrer I, Nicoll C~A, Ritchie D~A and Shields A~J
  2010 {\em Nature\/} {\bf 465} 594--597

\bibitem{Aharonovich2016}
Aharonovich I, Englund D and Toth M 2016 {\em Nat. Photon.\/} {\bf 10} 631--641

\bibitem{Fafard1999}
Fafard S, Wasilewski Z~R, Allen C~N, Picard D, Spanner M, McCaffrey J~P and
  Piva P~G 1999 {\em Phys. Rev. B\/} {\bf 59} 15368--15373

\bibitem{Wang2006b}
Wang L, Rastelli A and Schmidt O~G 2006 {\em J. Appl. Phys.\/} {\bf 100} 064313

\bibitem{Schneider2011}
Schneider C 2011 {\em Konzepte zur skalierbaren Realisierung von effizienten,
  halbleiterbasierten Einzelphotonenquellen\/} Ph.D. thesis
  Julius-Maximilians-Universit\"at W\"urzburg

\bibitem{Young2005}
Young R~J, Stevenson R~M, Shields A~J, Atkinson P, Cooper K, Ritchie D~A, Groom
  K~M, Tartakovskii A~I and Skolnick M~S 2005 {\em Phys. Rev. B\/} {\bf 72}
  113305

\bibitem{Ellis2007a}
Ellis D~J~P, Stevenson R~M, Young R~J, Shields A~J, Atkinson P and Ritchie D~A
  2007 {\em Appl. Phys. Lett.\/} {\bf 90} 011907

\bibitem{Dousse2010}
Dousse A, Suffczynski J, Beveratos A, Krebs O, Lemaitre A, Sagnes I, Bloch J,
  Voisin P and Senellart P 2010 {\em Nature\/} {\bf 466} 217--220

\bibitem{Kurtsiefer2000}
Kurtsiefer C, Mayer S, Zarda P and Weinfurter H 2000 {\em Phys. Rev. Lett.\/}
  {\bf 85}(2) 290--293

\bibitem{Aharonovich2014}
Aharonovich I and Neu E 2014 {\em Advanced Optical Materials\/} {\bf 2}
  911--928

\bibitem{Rogers2014}
Rogers L, Jahnke K, Teraji T, Marseglia L, M\"{u}ller C, Naydenov B, Schauffert
  H, Kranz C, Isoya J, McGuinness L and Jelezko F 2014 {\em Nat. Commun.\/}
  {\bf 5} 4739

\bibitem{Iwasaki2015}
Iwasaki T, Ishibashi F, Miyamoto Y, Doi Y, Kobayashi S, Miyazaki T, Tahara K,
  Jahnke K~D, Rogers L~J, Naydenov B, Jelezko F, Yamasaki S, Nagamachi S,
  Inubushi T, Mizuochi N and Hatano M 2015 {\em Sci. Rep.\/} {\bf 5} 12882

\bibitem{Castelletto2013}
Castelletto S, Johnson B~C, Iv{\'{a}}dy V, Stavrias N, Umeda T, Gali A and
  Ohshima T 2013 {\em Nature Materials\/} {\bf 13} 151--156

\bibitem{Choi2014}
Choi S, Johnson B~C, Castelletto S, Ton-That C, Phillips M~R and Aharonovich I
  2014 {\em Appl. Phys. Lett.\/} {\bf 104} 261101

\bibitem{Kolesov2012}
Kolesov R, Xia K, Reuter R, Stöhr R, Zappe A, Meijer J, Hemmer P and Wrachtrup
  J 2012 {\em Nat. Commun.\/} {\bf 3} 1029

\bibitem{Eichhammer2015}
Eichhammer E, Utikal T, Götzinger S and Sandoghdar V 2015 {\em New J. Phys.\/}
  {\bf 17} 083018

\bibitem{Mizuochi2012}
Mizuochi N, Makino T, Kato H, Takeuchi D, Ogura M, Okushi H, Nothaft M, Neumann
  P, Gali A, Jelezko F, Wrachtrup J and Yamasaki S 2012 {\em Nat Photon\/} {\bf
  6} 299--303

\bibitem{Lohrmann2015}
Lohrmann A, Iwamoto N, Bodrog Z, Castelletto S, Ohshima T, Karle T, Gali A,
  Prawer S, McCallum J and Johnson B 2015 {\em Nat. Commun.\/} {\bf 6} 7783

\bibitem{Hoegele2008}
H\"ogele A, Galland C, Winger M and Imamo{\u{g}}lu A 2008 {\em Phys. Rev.
  Letters\/} {\bf 100} 217401

\bibitem{Zhao2018}
Zhao S, Lavie J, Rondin L, Orcin-Chaix L, Diederichs C, Roussignol P,
  Chassagneux Y, Voisin C, Müllen K, Narita A, Campidelli S and Lauret J~S 2018
  {\em Nat. Commun.\/} {\bf 9} 3470

\bibitem{Wang2012}
Wang Q~H, Kalantar-Zadeh K, Kis A, Coleman J~N and Strano M~S 2012 {\em Nat.
  Nanotechnol.\/} {\bf 7} 699--712

\bibitem{Mak2016}
Mak K~F and Shan J 2016 {\em Nat. Photon.\/} {\bf 10} 216--226

\bibitem{Mak2010}
Mak K~F, Lee C, Hone J, Shan J and Heinz T~F 2010 {\em Phys. Rev. Letters\/}
  {\bf 105}

\bibitem{Splendiani2010}
Splendiani A, Sun L, Zhang Y, Li T, Kim J, Chim C~Y, Galli G and Wang F 2010
  {\em Nano Lett.\/} {\bf 10} 1271--1275

\bibitem{He2015}
He Y~M, Clark G, Schaibley J~R, He Y, Chen M~C, Wei Y~J, Ding X, Zhang Q, Yao
  W, Xu X, Lu C~Y and Pan J~W 2015 {\em Nat. Nanotechnol.\/} {\bf 10} 497--502

\bibitem{Srivastava2015}
Srivastava A, Sidler M, Allain A~V, Lembke D~S, Kis A and Imamo{\u{g}}lu A 2015
  {\em Nat. Nanotechnol.\/}

\bibitem{Chakraborty2015}
Chakraborty C, Kinnischtzke L, Goodfellow K~M, Beams R and Vamivakas A~N 2015
  {\em Nat. Nanotechnol.\/} {\bf 10} 507--511

\bibitem{Koperski2015}
Koperski M, Nogajewski K, Arora A, Cherkez V, Mallet P, Veuillen J~Y, Marcus J,
  Kossacki P and Potemski M 2015 {\em Nat. Nanotechnol.\/} {\bf 10} 503--506

\bibitem{You2015}
You Y, Zhang X~X, Berkelbach T~C, Hybertsen M~S, Reichman D~R and Heinz T~F
  2015 {\em Nat Phys\/} {\bf 11} 477--481

\bibitem{He2016a}
He Y~M, Iff O, Lundt N, Baumann V, Davanco M, Srinivasan K, H\"ofling S and
  Schneider C 2016 {\em Nat. Commun.\/} {\bf 7} 13409

\bibitem{Tran2015}
Tran T~T, Bray K, Ford M~J, Toth M and Aharonovich I 2015 {\em Nat.
  Nanotechnol.\/} {\bf 11} 37--41

\bibitem{Tran2016}
Tran T~T, Zachreson C, Berhane A~M, Bray K, Sandstrom R~G, Li L~H, Taniguchi T,
  Watanabe K, Aharonovich I and Toth M 2016 {\em Phys. Rev. Applied\/} {\bf 5}

\bibitem{Kianinia2017}
Kianinia M, Regan B, Tawfik S~A, Tran T~T, Ford M~J, Aharonovich I and Toth M
  2017 {\em {ACS} Photonics\/} {\bf 4} 768--773

\bibitem{Kumar2016}
Kumar S, Brot{\'{o}}ns-Gisbert M, Al-Khuzheyri R, Branny A, Ballesteros-Garcia
  G, S{\'{a}}nchez-Royo J~F and Gerardot B~D 2016 {\em Optica\/} {\bf 3} 882

\bibitem{Schneider2008}
Schneider C, Strau\ss M, S\"{u}nner T, Huggenberger A, Wiener D, Reitzenstein
  S, Kamp M, H\"{o}fling S and Forchel A 2008 {\em Appl. Phys. Lett.\/} {\bf
  92} 183101

\bibitem{Felici2009}
Felici M, Gallo P, Mohan A, Dwir B, Rudra A and Kapon E 2009 {\em Small\/} {\bf
  5} 938--943

\bibitem{Strittmatter2012}
Strittmatter A, Schliwa A, Schulze J~H, Germann T~D, Dreismann A, Hitzemann O,
  Stock E, Ostapenko I~A, Rodt S, Unrau W, Pohl U~W, Hoffmann A, Bimberg D and
  Haisler V 2012 {\em Appl. Phys. Lett.\/} {\bf 100} 093111

\bibitem{Heinrich2010}
Heinrich J, Huggenberger A, Heindel T, Reitzenstein S, H\"{o}fling S, Worschech
  L and Forchel A 2010 {\em Appl. Phys. Lett.\/} {\bf 96} 211117

\bibitem{Kern2016}
Kern J, Niehues I, Tonndorf P, Schmidt R, Wigger D, Schneider R, Stiehm T,
  de~Vasconcellos S~M, Reiter D~E, Kuhn T and Bratschitsch R 2016 {\em Advanced
  Materials\/} {\bf 28} 7101--7105

\bibitem{Branny2017}
Branny A, Kumar S, Proux R and Gerardot B~D 2017 {\em Nat. Commun.\/} {\bf 8}
  15053

\bibitem{Schmidt2007}
Schmidt O~G 2007 {\em Lateral Alignment of Epitaxial Quantum Dots\/} (Springer)

\bibitem{Hartmann1998}
Hartmann A, Ducommun Y, Loubies L, Leifer K and Kapon E 1998 {\em Appl. Phys.
  Lett.\/} {\bf 73} 2322--2324

\bibitem{Mereni2012}
Mereni L~O, Marquardt O, Juska G, Dimastrodonato V, O'Reilly E~P and Pelucchi E
  2012 {\em Phys. Rev. B\/} {\bf 85}(15) 155453

\bibitem{Verma2011}
Verma V~B, Stevens M~J, Silverman K~L, Dias N~L, Garg A, Coleman J~J and Mirin
  R~P 2011 {\em Opt. Express\/} {\bf 19} 4182--4187

\bibitem{Huggenberger2011a}
Huggenberger A, Schneider C, Drescher C, Heckelmann S, Heindel T, Reitzenstein
  S, Kamp M, H\"ofling S, Worschech L and Forchel A 2011 {\em J. Cryst.
  Growth.\/} {\bf 323} 194--197

\bibitem{Schneider2012a}
Schneider C, Huggenberger A, Gschrey M, Gold P, Rodt S, Forchel A, Reitzenstein
  S, H\"ofling S and Kamp M 2012 {\em Phys. Status Solidi A\/} {\bf 209}
  2379--2386

\bibitem{Albert2010}
Albert F, Stobbe S, Schneider C, Heindel T, Reitzenstein S, H\"{o}fling S,
  Lodahl P, Worschech L and Forchel A 2010 {\em Appl. Phys. Lett.\/} {\bf 96}
  151102 (pages~3)

\bibitem{Joens2013}
J\"{o}ns K~D K~D, Atkinson P, M\"{u}ller M, Heldmaier M, Ulrich S~M, Schmidt
  O~G and Michler P 2013 {\em Nano Lett.\/} {\bf 13} 126--130

\bibitem{Strittmatter2012a}
Strittmatter A, Holzbecher A, Schliwa A, Schulze J~H, Quandt D, Germann T~D,
  Dreismann A, Hitzemann O, Stock E, Ostapenko I~A, Rodt S, Unrau W, Pohl U~W,
  Hoffmann A, Bimberg D and Haisler V 2012 {\em Phys. Status Solidi A\/} {\bf
  209} 2411--2420

\bibitem{Ellis2007}
Ellis D~J~P, Bennett A~J, Shields A~J, Atkinson P and Ritchie D~A 2007 {\em
  Appl. Phys. Lett.\/} {\bf 90} 233514

\bibitem{Kumar2015}
Kumar S, Kaczmarczyk A and Gerardot B~D 2015 {\em Nano Lett.\/} {\bf 15}
  7567--7573

\bibitem{Palacios-Berraquero2017}
Palacios-Berraquero C, Kara D~M, Montblanch A~R~P, Barbone M, Latawiec P, Yoon
  D, Ott A~K, Loncar M, Ferrari A~C and Atat\"{u}re M 2017 {\em Nat. Commun.\/}
  {\bf 8} 15093

\bibitem{Shepard2017}
Shepard G~D, Ajayi O~A, Li X, Zhu X~Y, Hone J and Strauf S 2017 {\em 2D
  Materials\/} {\bf 4} 021019

\bibitem{Badolato2005}
Badolato A, Hennessy K, Atat\"ure M, Dreiser J, Hu E, Petroff P~M and
  Imamo\u{g}lu A 2005 {\em Science\/} {\bf 308} 1158--1161

\bibitem{Hennessy2007}
Hennessy K, Badolato A, Winger M, Gerace D, Atat\"{u}re M, Gulde S, F\"{a}lt S,
  Hu E~L and Imamo\u{g}lu A 2007 {\em Nature\/} {\bf 445} 896--899

\bibitem{Sapienza2015}
Sapienza L, Davan\c{c}o M, Badolato A and Srinivasan K 2015 {\em Nat.
  Commun.\/} {\bf 6} 7833

\bibitem{Purcell1946}
Purcell E~M 1946 {\em Proceedings of the American Physical Society\/} {\bf 69}
  681

\bibitem{Gerard1998}
Gérard J~M, Sermage B, Gayral B, Legrand B, Costard E and Thierry-Mieg V 1998
  {\em Phys. Rev. Letters\/} {\bf 81} 1110 -- 1113

\bibitem{Sapienza2017}
Sapienza L, Liu J, Song J~D, Fält S, Wegscheider W, Badolato A and Srinivasan K
  2017 {\em Scientific Reports\/} {\bf 7} 6205

\bibitem{Nogues2013}
Nogues G, Merotto Q, Bachelier G, Lee E~H and Song J~D 2013 {\em Appl. Phys.
  Lett.\/} {\bf 102} 231112

\bibitem{Kojima2013}
Kojima T, Kojima K, Asano T and Noda S 2013 {\em Appl. Phys. Lett.\/} {\bf 102}
  011110

\bibitem{Liu2017}
Liu J, Davanço M~I, Sapienza L, Konthasinghe K, Cardoso J~V~D~M, Song J~D,
  Badolato A and Srinivasan K 2017 {\em Rev. Sci. Instrum.\/} {\bf 88} 023116

\bibitem{Donatini2010}
Donatini F and Dang L~S 2010 {\em Nanotechnology\/} {\bf 21} 375303

\bibitem{Dousse2008}
Dousse A, Lanco L, Suffczy\'{n}ski J, Semenova E, Miard A, Lema\^{i}tre A,
  Sagnes I, Roblin C, Bloch J and Senellart P 2008 {\em Phys. Rev. Lett.\/}
  {\bf 101}(26) 267404

\bibitem{Gschrey2013}
Gschrey M, Gericke F, Sch\"{u}ßler A, Schmidt R, Schulze J~H, Heindel T, Rodt
  S, Strittmatter A and Reitzenstein S 2013 {\em Appl. Phys. Lett.\/} {\bf 102}
  251113

\bibitem{Lesik2013}
Lesik M, Spinicelli P, Pezzagna S, Happel P, Jacques V, Salord O, Rasser B,
  Delobbe A, Sudraud P, Tallaire A, Meijer J and Roch J 2013 {\em Phys. Status
  Solidi A\/} {\bf 210} 2055--2059

\bibitem{Schroeder2017}
Schr\"oder T, Trusheim M~E, Walsh M, Li L, Zheng J, Schukraft M, Sipahigil A,
  Evans R~E, Sukachev D~D, Nguyen C~T, Pacheco J~L, Camacho R~M, Bielejec E~S,
  Lukin M~D and Englund D 2017 {\em Nat. Commun.\/} {\bf 8} 15376

\bibitem{Yacobi1986}
Yacobi B~G and Holt D~B 1986 {\em J. Appl. Phys.\/} {\bf 59} R1--R24

\bibitem{Gschrey2015}
Gschrey M, Schmidt R, Schulze J~H, Strittmatter A, Rodt S and Reitzenstein S
  2015 {\em J. Vac. Sci. Technol. B\/} {\bf 33} 021603

\bibitem{Meijer2005}
Meijer J, Burchard B, Domhan M, Wittmann C, Gaebel T, Popa I, Jelezko F and
  Wrachtrup J 2005 {\em Appl. Phys. Lett.\/} {\bf 87} 261909

\bibitem{Tamura2014}
Tamura S, Koike G, Komatsubara A, Teraji T, Onoda S, McGuinness L~P, Rogers L,
  Naydenov B, Wu E, Yan L, Jelezko F, Ohshima T, Isoya J, Shinada T and Tanii T
  2014 {\em Applied Physics Express\/} {\bf 7} 115201

\bibitem{Radulaski2017}
Radulaski M, Widmann M, Niethammer M, Zhang J~L, Lee S~Y, Rendler T, Lagoudakis
  K~G, Son N~T, Janz{\'{e}}n E, Ohshima T, Wrachtrup J and Vu{\v{c}}kovi{\'{c}}
  J 2017 {\em Nano Lett.\/} {\bf 17} 1782--1786

\bibitem{Zhou2018}
Zhou Y, Mu Z, Adamo G, Bauerdick S, Rudzinski A, Aharonovich I and bo~Gao W
  2018 {\em New J. Phys.\/} {\bf 20} 125004

\bibitem{Schukraft2016}
Schukraft M, Zheng J, Schröder T, Mouradian S~L, Walsh M, Trusheim M~E, Bakhru
  H and Englund D~R 2016 {\em APL Photonics\/} {\bf 1} 020801

\bibitem{Junno1995}
Junno T, Deppert K, Montelius L and Samuelson L 1995 {\em Appl. Phys. Lett.\/}
  {\bf 66} 3627--3629

\bibitem{Toset2006}
Toset J, Casuso I, Samitier J and Gomila G 2006 {\em Nanotechnology\/} {\bf 18}
  015503

\bibitem{Sar2009}
van~der Sar T, Heeres E~C, Dmochowski G~M, de~Lange G, Robledo L, Oosterkamp
  T~H and Hanson R 2009 {\em Appl. Phys. Lett.\/} {\bf 94} 173104

\bibitem{Ampem-Lassen2009}
Ampem-Lassen E, Simpson D~A, Gibson B~C, Trpkovski S, Hossain F~M, Huntington
  S~T, Ganesan K, Hollenberg L~C~L and Prawer S 2009 {\em Opt. Express\/} {\bf
  17} 11287

\bibitem{Davanco2017}
Davanco M, Liu J, Sapienza L, Zhang C~Z, De~Miranda~Cardoso J~V, Verma V, Mirin
  R, Nam S~W, Liu L and Srinivasan K 2017 {\em Nat. Commun.\/} {\bf 8} 889

\bibitem{Katsumi2018}
Katsumi R, Ota Y, Kakuda M, Iwamoto S and Arakawa Y 2018 {\em Optica\/} {\bf 5}
  691

\bibitem{Reimer2012}
Reimer M~E, Bulgarini G, Akopian N, Hocevar M, Bavinck M~B, Verheijen M~A,
  Bakkers E~P, Kouwenhoven L~P and Zwiller V 2012 {\em Nat Commun\/} {\bf 3}
  737

\bibitem{Schell2011}
Schell A~W, Kewes G, Schr\"oder T, Wolters J, Aichele T and Benson O 2011 {\em
  Rev. Sci. Instrum.\/} {\bf 82} 073709

\bibitem{Wolters2010}
Wolters J, Schell A~W, Kewes G, N\"{u}sse N, Schoengen M, D\"{o}scher H,
  Hannappel T, L\"{o}chel B, Barth M and Benson O 2010 {\em Appl. Phys.
  Lett.\/} {\bf 97} 141108

\bibitem{Schroeder2011}
Schr\"{o}der T, Schell A~W, Kewes G, Aichele T and Benson O 2011 {\em Nano
  Lett.\/} {\bf 11} 198--202

\bibitem{Liebermeister2014}
Liebermeister L, Petersen F, v~Münchow A, Burchardt D, Hermelbracht J, Tashima
  T, Schell A~W, Benson O, Meinhardt T, Krueger A, Stiebeiner A, Rauschenbeutel
  A, Weinfurter H and Weber M 2014 {\em Appl. Phys. Lett.\/} {\bf 104} 031101

\bibitem{Zadeh2016}
Zadeh I~E, Elshaari A~W, J\"{o}ns K~D, Fognini A, Dalacu D, Poole P~J, Reimer
  M~E and Zwiller V 2016 {\em Nano Lett.\/} {\bf 16} 2289--2294

\bibitem{Schnauber2019}
Schnauber P, Singh A, Schall J, Park S~I, Song J~D, Rodt S, Srinivasan K,
  Reitzenstein S and Davanco M 2019 {\em Nano Lett.\/}

\bibitem{Sartison2017}
Sartison M, Portalupi S~L, Gissibl T, Jetter M, Giessen H and Michler P 2017
  {\em Sci. Rep.\/} {\bf 7} 39916

\bibitem{Strauf2006}
Strauf S, Hennessy K, Rakher M, Choi Y~S, Badolato A, Andreani L, Hu E, Petroff
  P and Bouwmeester D 2006 {\em Phys. Rev. Letters\/} {\bf 96} 127404

\bibitem{Claudon2010}
Claudon J, Bleuse J, Malik N~S, Bazin M, Jaffrennou P, Gregersen N, Sauvan C,
  Lalanne P and Gérard J~M 2010 {\em Nat. Photon.\/} {\bf 4} 174--177

\bibitem{Glauber1963}
Glauber R~J 1963 {\em Phys. Rev.\/} {\bf 130}(6) 2529--2539

\bibitem{Hong1987}
Hong C~K, Ou Z~Y and Mandel L 1987 {\em Phys. Rev. Lett.\/} {\bf 59}(18)
  2044--2046

\bibitem{Santori2002b}
Santori C, Fattal D, Vu\v{c}kovi\'{c} J, Solomon G~S and Yamamoto Y 2002 {\em
  Nature\/} {\bf 419} 594--597

\bibitem{DiGiuseppe2003}
Di~Giuseppe G, Atat\"{u}re M, Shaw M~D, Sergienko A~V, Saleh B~E~A, Teich M~C,
  Miller A~J, Nam S~W and Martinis J 2003 {\em Phys. Rev. A\/} {\bf 68} 063817

\bibitem{Helversen2019}
von Helversen M, B\"{o}hm J, Schmidt M, Gschrey M, Schulze J~H, Strittmatter A,
  Rodt S, Beyer J, Heindel T and Reitzenstein S 2019 {\em New J. Phys.\/} {\bf
  21} 035007

\bibitem{Prilmueller2018}
Prilm\"{u}ller M, Huber T, M\"{u}ller M, Michler P, Weihs G and
  Predojevi{\'{c}} A 2018 {\em Physical Review Letters\/} {\bf 121}

\bibitem{James2001}
James D~F~V, Kwiat P~G, Munro W~J and White A~G 2001 {\em Phys. Rev. A\/} {\bf
  64} 052312

\bibitem{Barnes2002}
Barnes W, Bj\"ork G, G\'erard J, Jonsson P, Wasey J, Worthing P and Zwiller V
  2002 {\em Eur. Phys. J. D\/} {\bf 18} 197--210

\bibitem{Reitzenstein2010}
Reitzenstein S and Forchel A 2010 {\em J. Phys. D\/} {\bf 43} 033001

\bibitem{Reithmaier2004}
Reithmaier J~P, S\c{e}k G, L\"offler A, Hofmann C, Kuhn S, Reitzenstein S,
  Keldysh L~V, Kulakovskii V~D, Reinecke T~L and Forchel A 2004 {\em Nature\/}
  {\bf 432} 197--200

\bibitem{Heindel2012}
Heindel T, Kessler C~A, Rau M, Schneider C, F\"urst M, Hargart F, Schulz W~M,
  Eichfelder M, Roßbach R, Nauerth S, Lermer M, Weier H, Jetter M, Kamp M,
  Reitzenstein S, H\"ofling S, Michler P, Weinfurter H and Forchel A 2012 {\em
  New J. Phys.\/} {\bf 14} 083001

\bibitem{Wang2017}
Wang H, He Y, Li Y~H, Su Z~E, Li B, Huang H~L, Ding X, Chen M~C, Liu C, Qin J
  and et~al 2017 {\em Nat. Photon.\/}  1749--4893

\bibitem{Schneider2009}
Schneider C, Heindel T, Huggenberger A, Weinmann P, Kistner C, Kamp M,
  Reitzenstein S, H\"{o}fling S and Forchel A 2009 {\em Appl. Phys. Lett.\/}
  {\bf 94} 111111

\bibitem{Somaschi2016}
Somaschi N, Giesz V, De~Santis L, Loredo J~C, Almeida M~P, Hornecker G,
  Portalupi S~L, Grange T, Ant\'{o}n C, Demory J, G\'{o}mez C, Sagnes I,
  Lanzillotti-Kimura N~D, Lema\'{i}tre A, Auffeves A, White A~G, Lanco L and
  Senellart P 2016 {\em Nat. Photon.\/} {\bf 10} 340--345

\bibitem{Gazzano2013}
Gazzano O, Michaelis~de Vasconcellos S, Arnold C, Nowak A, Galopin E, Sagnes I,
  Lanco L, Lema\^{i}tre A and Senellart P 2013 {\em Nat. Commun.\/} {\bf 4}
  1425

\bibitem{Nowak2014}
Nowak A~K, Portalupi S~L, Giesz V, Gazzano O, Dal~Savio C, Braun P~F, Karrai K,
  Arnold C, Lanco L, Sagnes I, Lema\^{i}tre Atre A and Senellart P 2014 {\em
  Nat. Commun.\/} {\bf 5} 3240

\bibitem{Unsleber2016}
Unsleber S, He Y~M, Gerhardt S, Maier S, Lu C~Y, Pan J~W, Gregersen N, Kamp M,
  Schneider C and H\"ofling S 2016 {\em Opt. Express\/} {\bf 24} 8539

\bibitem{Muller2007}
Muller A, Flagg E~B, Bianucci P, Wang X~Y, Deppe D~G, Ma W, Zhang J, Salamo
  G~J, Xiao M and Shih C~K 2007 {\em Phys. Rev. Lett.\/} {\bf 99}(18) 187402

\bibitem{Holmes2014}
Holmes M~J, Choi K, Kako S, Arita M and Arakawa Y 2014 {\em Nano Lett.\/} {\bf
  14} 982--986

\bibitem{Hadden2010}
Hadden J~P, Harrison J~P, Stanley-Clarke A~C, Marseglia L, Ho Y~L~D, Patton
  B~R, O´Brien J~L and Rarity J~G 2010 {\em Appl. Phys. Lett.\/} {\bf 97}
  241901

\bibitem{Siyushev2010}
Siyushev P, Kaiser F, Jacques V, Gerhardt I, Bischof S, Fedder H, Dodson J,
  Markham M, Twitchen D, Jelezko F and Wrachtrup J 2010 {\em Appl. Phys.
  Lett.\/} {\bf 97} 241902

\bibitem{Ates2012a}
Ates S, Sapienza L, Davanco M, Badolato A and Srinivasan K 2012 {\em Selected
  Topics in Quantum Electronics, IEEE Journal of\/} {\bf 18} 1711 --1721

\bibitem{Trojak2017}
Trojak O~J, Park S~I, Song J~D and Sapienza L 2017 {\em Appl. Phys. Lett.\/}
  {\bf 111} 021109

\bibitem{Trojak2018}
Trojak O~J, Woodhead C, Park S~I, Song J~D, Young R~J and Sapienza L 2018 {\em
  Appl. Phys. Lett.\/} {\bf 112} 221102

\bibitem{Gschrey2015b}
Gschrey M, Thoma A, Schnauber P, Seifried M, Schmidt R, Wohlfeil B, Kr\"{u}ger
  L, Schulze J~H, Heindel T, Burger S, Schmidt F, Strittmatter A, Rodt S and
  Reitzenstein S 2015 {\em Nat. Commun.\/} {\bf 6} 7662

\bibitem{Heindel2017b}
Heindel T, Rodt S and Reitzenstein S 2017  (Cham: Springer International
  Publishing) chap Single-Photon Sources Based on Deterministic Quantum-Dot
  Microlenses, pp 199--232

\bibitem{Thoma2016}
Thoma A, Schnauber P, Gschrey M, Seifried M, Wolters J, Schulze J~H,
  Strittmatter A, Rodt S, Carmele A, Knorr A, Heindel T and Reitzenstein S 2016
  {\em Phys. Rev. Letters\/} {\bf 116} 033601

\bibitem{Schnauber2016}
Schnauber P, Thoma A, Heine C~V, Schlehahn A, Gantz L, Gschrey M, Schmidt R,
  Hopfmann C, Wohlfeil B, Schulze J~H, Strittmatter A, Heindel T, Rodt S,
  Woggon U, Gershoni D and Reitzenstein S 2016 {\em Technologies\/} {\bf 4} 1

\bibitem{Schlehahn2015a}
Schlehahn A, Gaafar M, Vaupel M, Gschrey M, Schnauber P, Schulze J~H, Rodt S,
  Strittmatter A, Stolz W, Rahimi-Iman A, Heindel T, Koch M and Reitzenstein S
  2015 {\em Appl. Phys. Lett.\/} {\bf 107} 041105

\bibitem{Dusanowski2017}
Dusanowski {\L}, Holewa P, Mary{\'{n}}ski A, Musia{\l} A, Heuser T, Srocka N,
  Quandt D, Strittmatter A, Rodt S, Misiewicz J, Reitzenstein S and S\c{e}k G
  2017 {\em Opt. Express\/} {\bf 25} 31122

\bibitem{Sartison2018}
Sartison M, Engel L, Kolatschek S, Olbrich F, Nawrath C, Hepp S, Jetter M,
  Michler P and Portalupi S~L 2018 {\em Appl. Phys. Lett.\/} {\bf 113} 032103

\bibitem{Mueller2017}
M\"{u}ller M, Vural H, Schneider C, Rastelli A, Schmidt O~G, H\"{o}fling S and
  Michler P 2017 {\em Phys. Rev. Letters\/} {\bf 118}

\bibitem{Schwartz2016a}
Schwartz I, Cogan D, Schmidgall E~R, Don Y, Gantz L, Kenneth O, Lindner N~H and
  Gershoni D 2016 {\em Science\/} {\bf 354} 434--437

\bibitem{Stevenson2006}
Stevenson R~M, Young R~J, See P, Gevaux D~G, Cooper K, Atkinson P, Farrer I,
  Ritchie D~A and Shields A~J 2006 {\em Phys. Rev. B\/} {\bf 73} 033306

\bibitem{Avron2008}
Avron J~E, Bisker G, Gershoni D, Lindner N~H, Meirom E~A and Warburton R~J 2008
  {\em Phys. Rev. Letters\/} {\bf 100}

\bibitem{Juska2013}
Juska G, Dimastrodonato V, Mereni L~O, Gocalinska A and Pelucchi E 2013 {\em
  Nat Photon\/} {\bf 7} 527--531

\bibitem{Liu2019}
Liu J, Su R, Wei Y, Yao B, da~Silva S~F~C, Yu Y, Iles-Smith J, Srinivasan K,
  Rastelli A, Li J and Wang X 2019 {\em Nature Nanotechnology\/} {\bf 14}
  586--593

\bibitem{Bayer1998}
Bayer M, Gutbrod T, Reithmaier J~P, Forchel A, Reinecke T~L, Knipp P~A, Dremin
  A~A and Kulakovskii V~D 1998 {\em Phys. Rev. Lett.\/} {\bf 81} 2582--2585

\bibitem{Braun2013}
Braun T, Unsleber S, Baumann V, Gschrey M, Rodt S, Reitzenstein S, Schneider C,
  H\"ofling S and Kamp M 2013 {\em Appl. Phys. Lett.\/} {\bf 103} 191113

\bibitem{Mohan2010}
Mohan A, Felici M, Gallo P, Dwir B, Rudra A, Faist J and Kapon E 2010 {\em Nat.
  Photon.\/} {\bf 4} 302--306

\bibitem{Mohan2012}
Mohan A, Felici M, Gallo P, Dwir B, Rudra A, Faist J and Kapon E 2012 {\em Nat.
  Photon.\/} {\bf 6} 793--793

\bibitem{Mueller2014}
M\"{u}ller M, Bounouar S, J\"{o}ns K~D, Gl\"{a}\ss~l M and Michler P 2014 {\em
  Nat. Photon.\/} {\bf 8} 224--228

\bibitem{Wang2019}
Wang H, Hu H, Chung T~H, Qin J, Yang X, Li J~P, Liu R~Z, Zhong H~S, He Y~M,
  Ding X, Deng Y~H, Dai Q, Huo Y~H, H\"{o}fling S, Lu C~Y and Pan J~W 2019 {\em
  Phys. Rev. Lett.\/} {\bf 122} 113602

\bibitem{Wang2019a}
Wang H, He Y~M, Chung T~H, Hu H, Yu Y, Chen S, Ding X, Chen M~C, Qin J, Yang X,
  Liu R~Z, Duan Z~C, Li J~P, Gerhardt S, Winkler K, Jurkat J, Wang L~J,
  Gregersen N, Huo Y~H, Dai Q, Yu S, H\"{o}fling S, Lu C~Y and Pan J~W 2019
  {\em Nature Photonics\/} {\bf 13} 770--775

\bibitem{Stevenson2008}
Stevenson R~M, Hudson A~J, Bennett A~J, Young R~J, Nicoll C~A, Ritchie D~A and
  Shields A~J 2008 {\em Phys. Rev. Lett.\/} {\bf 101}(17) 170501

\bibitem{Huber2014}
Huber T, Predojevi\'{c} A, Khoshnegar M, Dalacu D, Poole P~J, Majedi H and
  Weihs G 2014 {\em Nano Lett.\/} {\bf 14} 7107--7114

\bibitem{Bounouar2018}
Bounouar S, de~la Haye C, Strau\ss M, Schnauber P, Thoma A, Gschrey M, Schulze
  J~H, Strittmatter A, Rodt S and Reitzenstein S 2018 {\em Appl. Phys. Lett.\/}
  {\bf 112} 153107

\bibitem{Winik2017}
Winik R, Cogan D, Don Y, Schwartz I, Gantz L, Schmidgall E~R, Livneh N,
  Rapaport R, Buks E and Gershoni D 2017 {\em Phys. Rev. B\/} {\bf 95} 235435

\bibitem{Khoshnegar2017}
Khoshnegar M, Huber T, Predojevi\'c A, Dalacu D, Prilm\"{u}ller M, Lapointe J,
  Wu X, Tamarat P, Lounis B, Poole P, Weihs G and Majedi H 2017 {\em Nat.
  Commun.\/} {\bf 8} 15716

\bibitem{Schmidgall2014a}
Schmidgall E~R, Schwartz I, Gantz L, Cogan D, Raindel S and Gershoni D 2014
  {\em Phys. Rev. B\/} {\bf 90} 241411(R)

\bibitem{Arashida2011}
Arashida Y, Ogawa Y and Minami F 2011 {\em Phys. Rev. B\/} {\bf 84}(12) 125309

\bibitem{Baier2004}
Baier M~H, Constantin C, Pelucchi E and Kapon E 2004 {\em Appl. Phys. Lett.\/}
  {\bf 84} 1967--1969

\bibitem{Mehta2010}
Mehta M, Reuter D, Wieck A~D, de~Vasconcellos S~M, Zrenner A and Meier C 2010
  {\em Appl. Phys. Lett.\/} {\bf 97} 143101

\bibitem{Schneider2012}
Schneider C, Heindel T, Huggenberger A, Niederstrasser T~A, Reitzenstein S,
  Forchel A, H\"{o}fling S and Kamp M 2012 {\em Appl. Phys. Lett.\/} {\bf 100}
  091108

\bibitem{Unrau2012}
Unrau W, Quandt D, Schulze J~H, Heindel T, Germann T~D, Hitzemann O,
  Strittmatter A, Reitzenstein S, Pohl U~W and Bimberg D 2012 {\em Appl. Phys.
  Lett.\/} {\bf 101} 211119

\bibitem{Sartison2019}
Sartison M, Seyfferle S, Kolatschek S, Hepp S, Jetter M, Michler P and
  Portalupi S~L 2019 {\em Appl. Phys. Lett.\/} {\bf 114} 222101

\bibitem{Chung2016}
Chung T~H, Juska G, Moroni S~T, Pescaglini A, Gocalinska A and Pelucchi E 2016
  {\em Nat. Photon.\/} {\bf 10} 782–--787

\bibitem{Xu2008}
Xu X, Brossard F, Hammura K, Williams D~A, Alloing B, Li L~H and Fiore A 2008
  {\em Appl. Phys. Lett.\/} {\bf 93} 021124

\bibitem{Haupt2010}
Haupt F, Oemrawsingh S~S~R, Thon S~M, Kim H, Kleckner D, Ding D, Suntrup D~J,
  Petroff P~M and Bouwmeester D 2010 {\em Appl. Phys. Lett.\/} {\bf 97} 131113

\bibitem{Cadeddu2016}
Cadeddu D, Teissier J, Braakman F~R, Gregersen N, Stepanov P, G\'{e}rard J~M,
  Claudon J, Warburton R~J, Poggio M and Munsch M 2016 {\em Appl. Phys.
  Lett.\/} {\bf 108} 011112

\bibitem{Schlehahn2018}
Schlehahn A, Fischbach S, Schmidt R, Kaganskiy A, Strittmatter A, Rodt S,
  Heindel T and Reitzenstein S 2018 {\em Sci. Rep.\/} {\bf 8} 1340

\bibitem{Munsch2013}
Munsch M, Malik N~S, Dupuy E, Delga A, Bleuse J, G\'erard J~M, Claudon J,
  Gregersen N and M\o{}rk J 2013 {\em Phys. Rev. Lett.\/} {\bf 110}(17) 177402

\bibitem{Veprik2005}
Veprik A, Riabzev S, Vilenchik G and Pundak N 2005 {\em Cryogenics\/} {\bf 45}
  117 -- 122

\bibitem{Schlehahn2015}
Schlehahn A, Kr\"uger L, Gschrey M, Schulze J~H, Rodt S, Strittmatter A,
  Heindel T and Reitzenstein S 2015 {\em Rev. Sci. Instrum.\/} {\bf 86} 013113

\bibitem{Snijders2018}
Snijders H, Frey J~A, Norman J, Post V~P, Gossard A~C, Bowers J~E, van Exter
  M~P, L\"offler W and Bouwmeester D 2018 {\em Phys. Rev. Applied\/} {\bf 9}(3)
  031002

\bibitem{Rickert2019}
Rickert L, Kupko T, Rodt S, Reitzenstein S and Heindel T 2019 {\em arXiv\/}
  1908.08408

\bibitem{He2019}
He Y~M, Wang H, Wang C, Chen M~C, Ding X, Qin J, Duan Z~C, Chen S, Li J~P, Liu
  R~Z, Schneider C, Atat\"{u}re M, H\"{o}fling S, Lu C~Y and Pan J~W 2019 {\em
  Nature Physics\/} {\bf 15} 941--946

\bibitem{Waks2002}
Waks E, Inoue K, Santori C, Fattal D, Vuckovic J, Solomon G~S and Yamamoto Y
  2002 {\em Nature\/} {\bf 420} 762--762

\bibitem{Intallura2007}
Intallura P~M, Ward M~B, Karimov O~Z, Yuan Z~L, See P, Shields A~J, Atkinson P
  and Ritchie D~A 2007 {\em Appl. Phys. Lett.\/} {\bf 91} 161103

\bibitem{Takemoto2010}
Takemoto K, Nambu Y, Miyazawa T, Wakui K, Hirose S, Usuki T, Takatsu M,
  Yokoyama N, Yoshino K, Tomita A, Yorozu S, Sakuma Y and Arakawa Y 2010 {\em
  Appl. Phys. Express\/} {\bf 3} 092802

\bibitem{Collins2010}
Collins R~J, Clarke P~J, Fern\'{a}ndez V, Gordon K~J, Makhonin M~N, Timpson
  J~A, Tahraoui A, Hopkinson M, Fox A~M, Skolnick M~S and Buller G~S 2010 {\em
  J. Appl. Phys.\/} {\bf 107} 073102

\bibitem{Gao2012}
Gao W~B, Fallahi P, Togan E, Miguel-Sanchez J and Imamoglu A 2012 {\em
  Nature\/} {\bf 491} 426--430

\bibitem{DeGreve2012}
De~Greve K, Yu L, McMahon P~L, Pelc J~S, Natarajan C~M, Kim N~Y, Abe E, Maier
  S, Schneider C, Kamp M, Hofling S, Hadfield R~H, Forchel A, Fejer M~M and
  Yamamoto Y 2012 {\em Nature\/} {\bf 491} 421--425

\bibitem{Delteil2015}
Delteil A, Sun Z, Gao W~b, Togan E, Faelt S and Imamoglu A 2015 {\em Nature
  Physics\/} {\bf 12} 218--223

\bibitem{Beveratos2002}
Beveratos A, Brouri R, Gacoin T, Villing A, Poizat J~P and Grangier P 2002 {\em
  Phys. Rev. Lett.\/} {\bf 89} 187901

\bibitem{Alleaume2004}
All\'eaume R, Treussart F, Messin G, Dumeige Y, Roch J~F, Beveratos A,
  Brouri-Tualle R, Poizat J~P and Grangier P 2004 {\em New J. Phys.\/} {\bf 6}
  92

\bibitem{Leifgen2014}
Leifgen M, Schröder T, Gädeke F, Riemann R, Métillon V, Neu E, Hepp C, Arend
  C, Becher C, Lauritsen K and Benson O 2014 {\em New J. Phys.\/} {\bf 16}
  023021

\bibitem{Hensen2015}
Hensen B, Bernien H, Dr{\'{e}}au A~E, Reiserer A, Kalb N, Blok M~S, Ruitenberg
  J, Vermeulen R~F~L, Schouten R~N, Abell{\'{a}}n C, Amaya W, Pruneri V,
  Mitchell M~W, Markham M, Twitchen D~J, Elkouss D, Wehner S, Taminiau T~H and
  Hanson R 2015 {\em Nature\/} {\bf 526} 682--686

\bibitem{Loredo2017}
Loredo J~C, Broome M~A, Hilaire P, Gazzano O, Sagnes I, Lemaitre A, Almeida
  M~P, Senellart P and White A~G 2017 {\em Phys. Rev. Lett.\/} {\bf 118}(13)
  130503

\bibitem{Giesz2015}
Giesz V, Portalupi S~L, Grange T, Ant\'on C, De~Santis L, Demory J, Somaschi N,
  Sagnes I, Lema\^{i}tre A, Lanco L and et~al 2015 {\em Phys. Rev. B\/} {\bf
  92} 161302(R)

\bibitem{Lo2012}
Lo H~K, Curty M and Qi B 2012 {\em Phys. Rev. Lett.\/} {\bf 108} 130503

\bibitem{Prtljaga2014}
Prtljaga N, Coles R~J, O'Hara J, Royall B, Clarke E, Fox A~M and Skolnick M~S
  2014 {\em Appl. Phys. Lett.\/} {\bf 104} 231107

\bibitem{Rengstl2015}
Rengstl U, Schwartz M, Herzog T, Hargart F, Paul M, Portalupi S~L, Jetter M and
  Michler P 2015 {\em Appl. Phys. Lett.\/} {\bf 107} 021101

\bibitem{Schnauber2018}
Schnauber P, Schall J, Bounouar S, H\"ohne T, Park S~I, Ryu G~H, Heindel T,
  Burger S, Song J~D, Rodt S and Reitzenstein S 2018 {\em Nano Lett.\/} {\bf
  18} 2336--2342

\end{thebibliography}
\bibliographystyle{iopart-num}

\end{document}